\documentclass[%
 reprint,
 amsmath,amssymb,
 aps,
]{revtex4-2}

\usepackage{graphicx}
\usepackage{dcolumn}
\usepackage{bm}

\usepackage[dvipsnames]{xcolor}

\newcommand{\mean}[1]{\langle{#1}\rangle}
\newcommand{\del}[0]{\partial}

\newcommand\ex{\textrm{e}}
\newcommand\DD{\textrm{D}}
\newcommand\pp{\textrm{p}}
\newcommand\PP{\textrm{P}}
\newcommand\dd{\textrm{d}}
\newcommand\ii{\textrm{i}}

\begin{document}

\preprint{APS/123-QED}

\title{Confinement-induced Transitions of\\ Chiral Active Particles and Dumbbells}

\author{Mike Brandt}
\email[]{mike.brandt2@cea.fr}
\affiliation{Institute for Theoretical Physics, Heidelberg University, 69120 Heidelberg, Germany}
\affiliation{Present address: Service de Physique de l’Etat Condens\'{e}, CEA, CNRS Universit\'{e} Paris-Saclay, 91191 Gif-sur-Yvette, France}
\author{Falko Ziebert}
\email[]{ziebert@thphys.uni-heidelberg.de}
\affiliation{Institute for Theoretical Physics, Heidelberg University, 69120 Heidelberg, Germany}
\author{Carsten Littek}
\email[]{littek@thphys.uni-heidelberg.de}
\affiliation{Institute for Theoretical Physics, Heidelberg University, 69120 Heidelberg, Germany}

\date{\today}

\begin{abstract}
We present a coherent framework for the stochastic description of a chiral Active Brownian Particle (cABP) and a chiral Active Brownian Dumbbell (cABD) 
in a harmonic trap. 
On the one hand, we employ the Martin-Siggia-Rose path integral formalism, using the passive 
Brownian motion as a reference state around which activity is accounted for as a perturbation series in the self-propulsion speed. This method 
is suited best 
to calculate the exact time-dependent correlation functions. On the other hand, we use the Fokker-Planck equation, which works best to obtain the steady-state probability distribution. 
We discuss in detail the transitions from a trap centered distribution for low activity/strong confinement to a ring-shaped 
distribution for 
high activity/low confinement and
the effect of chirality, i.e.~the
active angular velocity. For the dumbbell, a series of such transitions occurs and the correlation functions yield information about the particle's conformation and orientation. This establishes the combined
framework as a powerful tool to investigate
stochastic active particle systems.
\end{abstract}

\maketitle

\section{Introduction}

The most studied realization of “active matter” \cite{Cates_whatisactivematter}
are self-propelled particles, i.e.~particles which are in an 
out-of-equilibrium state  using energy sources/fluxes to move autonomously.
While microswimmers like bacteria \cite{elgeti-Gompert-2015} are the most common biological realization,
many lab-built realizations are 
nowadays available, mostly falling
in the colloidal domain \cite{RevModPhys.88.045006}.
They hence
typically involve strong fluctuations,
calling for a stochastic description.

An interesting subclass of such
"active Brownian particles"
are those displaying chirality
\cite{van-teeffelen-2008,sevilla-2016}.
Examples range from 
artificial swimmers 
with shape anisotropy 
\cite{Kümmel2013}
or activity anisotropy \cite{EbbensBCS},
to biological objects like sperm cells 
spinning on circles due to their
flagellar beating patterns 
\cite{Friedrich2010,Elgeti_Sperm}, 
or crawling parasites
\cite{3D-Chiral-OU-Process}, where
chirality is caused by the
surface flow of adhesins that propels them \cite{NatPhysMalaria}.

Focusing here on the dynamics and
statistics of single active particles,
a situation that attracted much 
interest is the
interplay of activity and confining 
potentials
\cite{LöwenBechinger_trap,Militaru_pot}.
One hallmark of confinement 
is the occurrence of 
one of the simplest
nonequilibrium transitions, namely 
one from a trap-centered probability distribution 
to a "donut-shaped" distribution 
peaked at a finite 
radial distance
upon increasing activity/decreasing confinement. This transition was 
first theoretically predicted \cite{Holger_donut}, and
later on experimentally observed by
confining active colloids in an
acoustic trap \cite{TakatoriBrady}.

On the theory side, 
there is an increasing literature 
on active particles
trapped in potentials 
\cite{Holger_donut,Schehr_donut,
malakar-2020, PhysRevLett.129.158001,littek-2025}
and also confined chiral active particles 
have already been discussed \cite{jahanshahi-2017,Caprini_CS_in_pot,
littek-2025}.
Finally, models for more 
complex particles have been
introduced.  
Dumbbells, i.e.~two connected particles, were mostly used 
to implement weak shape anisotropy in simulations
\cite{LaeticiaPRE,LaeticiaPRL,Singh},
but active Brownian dumbbells \cite{Winkler_ABD}
assuming spring forces connecting the particles,
as well as longer active polymers
\cite{WinklerGompper}
up to effective models for active crystals \cite{Caprini_active_clusters} have been proposed.
Methods-wise, due to the inherent stochasticity,
mostly the Langevin approach 
\cite{van-teeffelen-2008,Holger_donut,Winkler_ABD} and the equivalent
Fokker-Planck equation approach (FPE)
\cite{sevilla-2016,malakar-2020, PhysRevLett.129.158001}
have been used.
The stochastic path integral, 
employing the Doi-Peliti approach \cite{pruessner_DoiPeliti},
the field theoretic version of the FPE
\cite{ZhangNJP,PruessnerRPR}
or the Martin-Siggia-Rose-Janssen-DeDominicis (MSRJD) formalism
\cite{Crisanti,SilvanoBarci,littek-2025},
is a mathematically equivalent  alternative under recent development
that offers some advantages.

Here we use a combination
of the MSRJD path integral formalism
and the Fokker-Planck equation
to thoroughly study the behaviour of
chiral active Brownian 
particles and 
dumbbells trapped in
a harmonic potential. This work is structured as follows:
first, in section \ref{sec:methods} we briefly describe
both methods and how they are applied to 
the stochastic equations of motion of active particles.
Then, in section \ref{sec: Brownian Circle Swimmer}
we study in detail the chiral Active Brownian Particle
(cABP, also referred to as
Brownian circle swimmer) with both methods.
As results we give the exact relevant moments 
and the steady-state probability distribution
as a series expansion, allowing to quantify  
its shape transition in the trap as a function
of velocity, angular velocity and trap strength.
In section \ref{dumbbell} we apply the same
procedure to the chiral Active Brownian Dumbbell (cABD).
Here, in addition we discuss interesting features
of the mean orientation of the dumbbell 
and describe a series of shape 
transitions in the trap, with respect to the 
distribution of the active particle and 
the passive cargo constituting the dumbbell.
Finally, in section \ref{sec:conclusions} we present our conclusions and outlook.

\section{Stochastic Path Integrals 
and Fokker-Planck Equation}
\label{sec:methods}

We begin by outlining the methods employed 
to describe the stochastic motion, 
the Martin-Siggia-Rose-Janssen-DeDominicis 
(MSRJD) path integral 
\cite{PhysRevA.8.423,Janssen,de-dominicis-1976}, 
and the Fokker-Planck equation (FPE)
\cite{Fokker1914,Planck17}. 
The discussion of the path integral approach 
builds on the detailed derivation 
presented in Ref.~\cite{littek-2025}. 
The Fokker-Planck equation is a standard tool in
stochastic processes \cite{risken-1985} and was 
already applied to the Active Brownian Particle 
in a harmonic trap, 
e.g.~in Refs.~\cite{malakar-2020,PhysRevLett.129.158001}. 
Both approaches are equivalent 
for describing stochastic dynamics, 
as discussed for instance in Ref.~\cite{zinn-justin-2002}. 

As we are going to point out, the path integral is best suited for the calculation of time-dependent moments without the need of the distribution function, while the time-dependent Fokker-Planck equation for the probability distribution is typically hard to solve. However, as the trapped systems considered here reach a steady-state at long times, the stationary Fokker-Planck equation allows the derivation of the steady-state 
distribution function via a rather simple recursion relation. 
In contrast, the evaluation of the perturbation series in activity in the path integral approach rapidly becomes difficult. 
 
\subsection{The MSRJD Path Integral}

We consider a system with $d$ degrees of freedom, represented by the state vector $X$, governed by the overdamped Langevin equation
\begin{equation}\label{eq: Langevin Equations}
    \begin{cases}
        \dot{X}_\mu(t) =F_\mu(X(t))+\sqrt{2D_\mu} \, \eta_\mu(t),\\
        X(0) =X_0,
    \end{cases}
\end{equation}
where $F_\mu$ denotes the deterministic drift and $D_\mu$ is the diffusion coefficient corresponding to the component $X_\mu$. The term $\eta_\mu(t)$ represents Gaussian white noise
\begin{equation}
    \langle\eta_\mu(t)\rangle=0 \quad \text{and} \quad \langle\eta_\mu(t)\eta_\nu(t')\rangle=\delta_{\mu\nu}\delta(t-t').
\end{equation}
A physical trajectory $X(X_0,\eta)$ is uniquely determined by a fixed initial condition $X(t=0)=X_0$ and a specific noise realization $\eta$. 

Following the Martin-Siggia-Rose formalism \cite{PhysRevA.8.423}, we enforce the Langevin equation of motion, $\mathrm{EoM}(X,X_0,\eta) \equiv \dot{X} - F(X) - \sqrt{2D}\,\eta = 0$, in the path integral via a functional Dirac delta distribution, which we represent as a path integral with an auxiliary response field $\chi$ according to 
\begin{equation}
    \delta\big(\mathrm{EoM}(X,X_0,\eta)\big) = \int \DD\chi \: \ex^{-\ii \left(\chi \big| \mathrm{EoM}(X,X_0,\eta) \right)},
\end{equation}
where we introduced the notation
\begin{equation}
    (X|Y)=\sum_{\mu=1}^d\int_0^\infty \dd  t \: X_\mu (t)Y_\mu(t).
\end{equation}
Averaging over the Gaussian noise $\eta$ and introducing source fields $h_\mu$ and $g_\mu$ for $X_\mu$ and $\chi_\mu$, respectively, yields the generating functional for a fixed initial condition $X(0)=X_0$,
\begin{equation}
        \mathcal{Z}[h,g]=\int_{X(0)=X_0}\DD \chi \DD X \, \ex^{-S(X,\chi) +(X|h)+(\chi|g)},
\end{equation}
where we call 
\begin{equation}
    S(X,\chi)= (\chi|D\chi)+\ii\left(\chi\big|\dot{X}-F(X)\right),
\end{equation}
which is a functional in $X$ and $\chi$,
the  MSRJD action of the system. 

\subsubsection{Perturbative Expansion in Activity}

We now consider specifically the Langevin equations 
for a two-dimensional (2D) chiral Active Brownian Particle (cABP) confined in a harmonic trap with stiffness $\alpha$. 
The degrees of freedom are the spatial position $\vec{q} = (q_x,q_y)$ and the orientation vector $\vec{n}(\vartheta) = (\cos\vartheta, \sin\vartheta)$, with the angle $\vartheta \in [0,2\pi)$. The particle propels itself with a constant speed $v_0$ along $\vec{n}$ and is subject to a constant angular velocity $\Omega$, which breaks chiral symmetry. Setting the friction coefficients to unity, the overdamped Langevin equations read
\begin{align}
    \dot{\vec{q}}(t) &= v_0\vec{n}(\vartheta) - \alpha \vec{q}(t) + \sqrt{2D_q}\,\vec{\eta}_q(t),\label{eq: Langevin for q}\\
    \dot{\vartheta}(t) &= \Omega + \sqrt{2D_\vartheta}\,\eta_\vartheta(t). \label{eq: Langevin for Vartheta}
\end{align}
with $D_q$ and $D_\vartheta$ characterizing the strengths of the translational and rotational Gaussian white noise.

Because of the nonlinear coupling introduced by the trigonometric functions in $\vec{n}(\vartheta)$, Eq.~\eqref{eq: Langevin for q} is analytically inaccessible for exact closed-form solutions. Consequently, we employ a perturbative approach in the self-propulsion speed $v_0$. To this end, we decompose the action into a free and an interacting (active) part:
\begin{equation}
    S(\vec{q},\vartheta,\chi) = S_0(\vec{q},\vartheta,\chi) + S_\mathrm{A}(\vec{q},\vartheta,\chi).
\end{equation}
The free action, $S_0$, describes two independent processes -- a two-dimensional Ornstein-Uhlenbeck process in the harmonic trap and a chiral rotational diffusion:
\begin{align}
    S_0(\vec{q},\vartheta,\chi) &= D_q(\vec{\chi}_q|\vec{\chi}_q) + \ii\big(\vec{\chi}_q\big|\dot{\vec{q}}+\alpha\vec{q}\big) \nonumber\\ 
    &\quad + D_\vartheta(\chi_\vartheta|\chi_\vartheta) + \ii\big(\chi_\vartheta\big|\dot{\vartheta} -\Omega\big).
\end{align}
The active part, $S_\mathrm{A}$, encodes the self-propulsion and couples the positional response field $\vec{\chi}_q$ to the orientation angle $\vartheta$
\begin{equation}\label{SA}
    S_\mathrm{A}(\vec{q},\vartheta,\chi) = -\ii v_0 \big(\vec{\chi}_q \big| \vec{n}(\vartheta)\big).
\end{equation}
Introducing source fields $h_\mu$ and $g_\mu$, the generating functional of the unperturbed system factorizes as
\begin{equation}
    \mathcal{Z}_0[h,g] = \mathcal{Z}^\alpha[\vec{h}_q,\vec{g}_q] \cdot \mathcal{Z}^\vartheta[h_\vartheta,g_\vartheta].
\end{equation}
Here, $\mathcal{Z}^\alpha$ and $\mathcal{Z}^\vartheta$ are the generating functionals for the 2D Ornstein-Uhlenbeck process and the 2D chiral rotational diffusion, respectively. 
For the initial condition $X_0=(\vec q_0,\vartheta_0)$ 
they evaluate to
\begin{widetext}
\begin{align}
    \mathcal{Z}^\alpha[\vec{h}_q,\vec{g}_q] &= \exp\left[ \big(\ex^{-\alpha t}\vec{q}_0\big|\vec{h}_q\big) + \frac{1}{2}\big(\vec{h}_q\big|C_\alpha\big|\vec{h}_q\big) + \big(\vec{h}_q\big|G_\alpha\big|\vec{g}_q\big) \right], \\
    \mathcal{Z}^\vartheta[h_\vartheta,g_\vartheta] &= \exp\left[\big(\vartheta_0|h_\vartheta\big) + \big(\Omega t|h_\vartheta\big) + \frac{1}{2}\big(h_\vartheta|C_\vartheta|h_\vartheta\big) + \big(h_\vartheta|G_\vartheta|g_\vartheta\big)\right]. 
\end{align}
\end{widetext}
Here, the propagators can be obtained by solving the corresponding Dyson-Schwinger equations, see Ref.~\cite{littek-2025}.
$G_\alpha$ and $C_\alpha$ govern 
the Ornstein-Uhlenbeck dynamics,
\begin{align}
    G_\alpha(t,t') &= -i\Theta(t-t')\ex^{-\alpha (t-t')},\\
    C_\alpha(t,t') &= \frac{D_q}{\alpha} \left[ \ex^{-\alpha|t-t'|} - \ex^{-\alpha(t+t')} \right]. \label{eq: C_alpha sym}
\end{align}
Similarly, $G_\vartheta$ and
$C_\vartheta$
encode the free angular diffusion:
\begin{align}
     G_\vartheta(t,t') &= -\ii\Theta(t-t'), \\
     C_\vartheta(t,t') &= 2D_\vartheta\min\{t,t'\}.
\end{align}

Having a closed expression for 
$\mathcal{Z}_0[h,g]$, we now have to deal with the activity. We elevate the fields in $S_\mathrm{A}$ to functional derivatives with respect to the source fields
\begin{equation}
    X_\mu(t) \rightarrow \frac{\delta}{\delta h_\mu(t)} \quad \text{and} \quad \chi_\mu(t) \rightarrow \frac{\delta}{\delta g_\mu(t)},
\end{equation}
for $\mu \in \{x,y,\vartheta\}$. This allows to pull the active part of the action in front of the free generating functional of the passive reference system
\begin{equation}\label{eq:full-gen-func}
    \mathcal{Z}[h,g] = \ex^{- S_\mathrm{A}\left(\frac{\delta}{\delta h},\frac{\delta}{\delta g}\right)} \mathcal{Z}_0[h,g].
\end{equation}
This expression describes deviations from the passive process by activity and can be systematically evaluated order by order in $v_0$ -- since the velocity enters $S_\mathrm{A}$ as a linear factor,
cf.~Eq.~(\ref{SA}) --
by Taylor expanding the exponential. 

Technically, this is done best by converting
the active part of the action to an operator, $S_\mathrm{A} \mapsto -i v_0\hat{A}$ with
\begin{equation}\label{eq: A-operator}
    \hat{A} = \frac{1}{2}\big\{\big(a^+|\phi^+\big) + \big(a^-|\phi^-\big)\big\}\,,
\end{equation}
where the operators $a^\sigma(t)$ and $\phi^\sigma(t)$ are given by
\begin{align}
     a^{\sigma}(t) &= \frac{\delta}{\delta g_x(t)} + \ii\sigma \frac{\delta}{\delta g_y(t)},\\
     \phi^{\sigma}(t) &= \ex^{- \ii\sigma\frac{\delta}{\delta h_{\vartheta}(t)}}
\end{align}
for $\sigma \in \{+,-\}$.
Applying these operators to the free generating functional $\mathcal{Z}_0[h,g]$ yields the rules
\begin{align}
    \phi^{\sigma}(t)\mathcal{Z}_0[h,g] &= \mathcal{Z}^{\alpha}[\vec{h}_q,\vec{g}_q] \, \mathcal{Z}^{\vartheta}[h_{\vartheta}-\ii\sigma\delta_t, g_{\vartheta}],\\
    a^{\sigma}(t)\mathcal{Z}_0[h,g] &= \left[ \big(\vec{h}_q|G_\alpha\big)_x(t) + \ii\sigma\big(\vec{h}_q| G_\alpha\big)_y(t) \right] \mathcal{Z}_0[h,g],
\end{align}
where $\delta_t$ with 
$\delta_t(t')=\delta(t' - t)$ denotes the shifted Dirac delta.

\subsubsection{Moments and probability distribution}

Correlation functions can now be calculated
by evaluating functional derivatives 
of $\mathcal{Z}[h,g]$ with respect to 
the source fields at $h,g=0$. 
Crucially, this operator formalism reveals 
a strict termination condition for the 
perturbation series: Since the action of 
$\hat{A}^N$ on $\mathcal{Z}_0$ generates
terms proportional to $|\vec{h}_q|^N$, setting $\vec{h}_q = 0$ annihilates any terms of higher order. Consequently, the perturbation series for the $N^{\mathrm{th}}$ moment of $\vec{q}$ terminates exactly at order $\mathcal{O}(v_0^N)$. This property allows to compute exact, non-perturbative results for the spatial correlation functions, 
e.g, the averaged position or the 
mean squared displacement (MSD). 

The spatial probability distribution in Fourier space can also be calculated from the generating functional by applying the exponential
\begin{align}
    \hat{\mathrm{p}}(t,\vec k) = \bigg(\ex^{-\ii\vec k \cdot \frac{\delta}{\delta \vec{h}_q}}\mathcal{Z}[h,g]\bigg)\Bigg|_{h,g=0}\equiv\mathcal{Z}[\vec h_q =-\ii\vec k\delta _t, 0]. \label{eq: General Probability Formula}
\end{align}
Underlying this expression is the Fourier representation of $\delta\left(\vec{x}-\frac{\delta}{\delta \vec{h}_q}\right)$,
needed to calculate 
$p(t,x)=\langle\delta(\vec{x}-\vec q(t))\rangle$ \cite{littek-2025}.
Importantly, because the exponential of a derivative acts like a shift,
unlike for finite moments where evaluation is at zero, one has to evaluate at the shifted source 
$\vec h_q=-i\vec k\delta_t$. 
Hence, applying the activity operator several times generates corrections 
that do not vanish. Consequently, 
the perturbation series for the 
probability distribution 
does not terminate at finite order in $v_0$
and has to be truncated to obtain an 
approximate solution. 

Due to the harmonic confinement considered here,
the probability distribution of the cABP
converges to a stationary distribution for long 
times. Evaluating this steady state via the 
path integral framework requires computing the 
generating functional at a given order 
in $v_0$ -- a task involving $N$ nested time 
integrals -- and subsequently taking the 
limit $t \to \infty$. While we did so
up to order $v_0^4$, there is a much simpler method
to obtain this steady state, namely the 
Fokker-Planck approach that is explained next.

\subsection{The Fokker Planck Equation}
\label{secFPE}

As an alternative, the stationary Fokker-Planck
equation (FPE) on the phase space $\mathbb{R}
^2\times[0,2\pi)$ provides a partial differential equation for the stationary distribution. 
As explained below, this reduces the problem 
to the algebraic task of solving a 
mode-coupling recursion.

The drift and diffusion coefficients of the Fokker-Planck equation can be directly
extracted from the corresponding Langevin dynamics \cite{risken-1985}. 
The translational drift is given by the harmonic restoring force $-\alpha\vec{x}$ and 
the self-propulsion $v_0\vec{n}(\vartheta)$, 
the angular drift is given by the angular velocity $\Omega$. Since we assumed additive noise independently for the two translational degrees of freedom and the angle, 
the diffusion tensor is diagonal with constants $D_q$ and $D_\vartheta$. The stationary Fokker-Planck equation for the probability density $\PP(\vec{x},\vartheta)$ therefore reads
\begin{align}\label{FPEfull}
    0 &= \alpha \nabla_{\vec{x}} \cdot (\vec{x}\,\PP) - v_0 \vec{n}(\vartheta) \cdot \nabla_{\vec{x}} \PP + D_q \nabla_{\vec{x}}^2 \PP \nonumber \\
    &\quad - \Omega \partial_\vartheta \PP + D_\vartheta \partial_\vartheta^2 \PP.
\end{align}
To analytically approach this partial differential equation, we perform a mode decomposition. First, we transform the spatial coordinates into Fourier space
\begin{equation}
    \hat\PP(\vec{k},\vartheta) = \int \dd^2 x \; \PP(\vec{x}, \vartheta) \ex^{-\ii\vec{k} \cdot \vec{x}}.
\end{equation}
Because the harmonic potential and the translational and rotational noises are isotropic, the stationary probability distribution must exhibit the same continuous rotational symmetry. Consequently, the stationary density in Fourier space can only depend on the wave-vector magnitude $k=|\vec{k}|$ and the relative angle between the orientation and the wave vector. In order to deal with the scalar product $\vec n(\vartheta)\cdot \vec k$ from the self-propulsion term in Fourier space, 
we write the wave vector in polar coordinates 
$\vec k = k(\cos \varphi, \sin \varphi)$ 
and perform an angular decomposition 
according to
\begin{equation}
    \hat\PP(\vec k,\vartheta)= \sum_{m\in\mathbb Z}\hat\PP_m(k)\ex^{-\ii m(\vartheta-\varphi)}.
\end{equation} 

Inserting these expansions into the stationary FPE, Eq.~(\ref{FPEfull}), yields
\begin{align}\label{eq: cABP equation in Fourier Space}
    0 &= -\alpha k\del_k\hat\PP_m + (D_q k^2 + D_\vartheta m^2 + \ii m\Omega)\hat\PP_m \nonumber\\ 
    & \quad +\frac{\ii v_0 k}{2}\big(\hat\PP_{m-1} + \hat\PP_{m+1}\big).
\end{align}
Similar as employed in the path integral approach, we now introduce a passive reference state for $v_0=0$ and use it to expand the full solution as a power series in $v_0$.

\subsubsection{Passive Reference State}

While $\hat\PP_m(k)$ denotes the $m^{\mathrm{th}}$ angular Fourier mode of the exact density, 
we now introduce $\hat\PP^{(n)}(\vec{k}, \vartheta)$ as the $n^{\mathrm{th}}$-order perturbative correction in the activity $v_0$. 
Setting $v_0=0$, the ordinary differential equation in $k$
\begin{equation}
    \alpha k \del_k \hat\PP_m^{(0)}=\big(D_qk^2+D_\vartheta m^2 + im\Omega\big)\hat\PP_m^{(0)}
\end{equation}
can be solved by separation of variables, yielding 
\begin{equation}
    \hat \PP_m^{(0)}(k) = \mathcal{N}_m \cdot k^{-\frac{D_\vartheta m^2+im\Omega}{\alpha}} \ex^{-\frac{D_q k^2}{2\alpha}}   
\end{equation}
with a normalization constant $\mathcal{N}_m$.
We must require that $m=0$, because otherwise the solution would be diverging at $k=0$, which is unphysical because it must equal the normalization.
The value of $\mathcal{N}_0$ follows from normalization of the marginal distribution 
over $\vartheta$
which simply leads to $\mathcal{N}_0=1/2\pi$. 
The passive reference state hence reads
\begin{equation}
    \hat \PP ^{(0)}_0(k)=\frac{1}{2\pi}\,\ex^{-\frac{
    D_q
    }{2\alpha}k^2},
\end{equation}
which could also have been guessed.
It is the product of a uniform distribution for the orientation angle $\vartheta$ and a Gaussian distribution of the passive Brownian particle 
in the harmonic potential.

\subsubsection{Power Series Expansion}

Now, we expand around the passive reference state by making the power series ansatz
\begin{align} \label{FPEPstatsol}
    \hat{\PP}_m(k)&=\bigg(\sum_{j=0}^\infty g_{m,j}k^j\bigg)\hat \PP_0^{(0)}(k). 
\end{align}
Note that in order to guarantee normalizability,  $j\geq 0$ must hold, and that one has
$g_{0,0}=1$ by consistency of notation.

Inserting the expansion into 
Eq.~(\ref{eq: cABP equation in Fourier Space}) leads to a recursion relation for the expansion coefficients
\begin{align}\label{eq:recursion-relation}
    g_{m,j}= \frac{-iv_0}{2(\alpha j + D_\vartheta m^2+im\Omega)}\big(g_{m-1,j-1}+g_{m+1, j-1}\big).
\end{align}
An important feature of this relation is that it couples coefficients of order $j$ to those of order $j-1$ via a factor proportional to $v_0$. In turn, this implies that $g_{m,j} \sim v_0^j$ and, consequently, the series expansion in $g_{m,j}$ constitutes a perturbative expansion 
in the self-propulsion velocity. 
This is completely analogous to the perturbative expansion of the path integral formalism.
Coefficients calculated by the recursion and by
the path integral formalism after 
taking the limit 
$t\to\infty$ are in fact identical, 
reflecting the mathematical equivalence between the two approaches \cite{zinn-justin-2002}.

The recursion relation is not solvable in  closed form. 
In fact, a solution for the general scalar recursion relation for $c_n$, with coefficients $Q_n^\pm\in \mathbb{R}$,
\begin{equation}
Q_nc_n+Q^+_nc_{n+1}+Q_n^{-}c_{n-1}=0
\end{equation}
can only be expressed in an infinitely long continued fraction \cite{risken-1985}
and we have to truncate the expansion at a certain order $N$. 
However, this approach reduces the problem of calculating the stationary probability distribution for the harmonically confined cABP to a linear recursion relation, 
instead of the $N!$-many integrals from the time-dependent formulation in the MSRJD-formalism.

\section{Chiral Active Brownian Particle}
\label{sec: Brownian Circle Swimmer}

We now use the methods introduced in Section \ref{sec:methods} to quantify the stochastic motion of a confined, chiral Active Brownian Particle (cABP). 
We first employ the generating functional, 
Eq.~(\ref{eq:full-gen-func}), to derive time-dependent expressions for correlation functions of both, velocity and position, and characterize the oscillating crossover at intermediate times. Here, the advantage of the MSRJD formalism is that the explicit form of the probability distribution is not necessary. We then present a perturbative expression for the time-dependent probability distribution $\PP(\vec q,t)$ up to second order in self-propulsion. 
Finally, we present the solution of the 
recursion relation of the FPE
and the resulting 
stationary probability distribution,
Eqs.~(\ref{FPEPstatsol})-(\ref{eq:recursion-relation}).
For strong harmonic confinement or low self-propulsion the distribution is unimodal at the center, 
while for weak harmonic confinement or high self-propulsion the distribution is radially peaked 
as the particle can climb the potential. We obtain the phase diagram harmonic stiffness $\alpha$ vs.~self-propulsion speed $v_0$ to characterize this shape transition
and show that angular velocity $\Omega$
shifts this transition to larger activity,
i.e.~hinders the transition to a radially peaked
distribution.

\subsection{Moments}
\label{sec:cABP-moments}

The stochastic motion of the cABP can be characterized by correlation functions 
of its velocity/orientation 
$\vec v=v_0\vec n(\vartheta)$ 
and position $\vec q$. 
Both can be derived from the generating functional, Eq.~(\ref{eq:full-gen-func}), by evaluating functional derivatives with respect to the sources $h_\mu(t)$ at $h,g=0$. 

The dynamics of the velocity is completely included in the passive reference system
and hence no perturbative expansion is needed.
Without loss of generality, we omit the self-propulsion speed $v_0$ and express correlations in terms of the orientation vector $\vec n$. We make use of the operator $\phi^\sigma(t)$ to express trigonometric functions
\begin{align}
    \cos(\vartheta(t))&=\left.\frac{1}{2}\left[\phi^-(t)+\phi^+(t)\right]\mathcal{Z}\right|_{h,g=0},\nonumber\\
    \sin(\vartheta(t))&=\left.\frac{1}{2\ii}\left[\phi^-(t)-\phi^+(t)\right]\mathcal{Z}\right|_{h,g=0}.
\end{align}
In this way, the mean orientation is simple 
to evaluate to
\begin{equation}
    \mean{\vec n(\vartheta(t))}=\ex^{-D_\vartheta t}\vec n(\vartheta_0+\Omega t).
\end{equation}
The mean orientation vector rotates with 
constant angular velocity $\Omega$ until it is reduced to zero by rotational diffusion. 
In other words, for long times rotational diffusion fully randomizes the orientation despite the particle's chirality.

\begin{figure*}
    \centering
    \includegraphics[width=\linewidth]{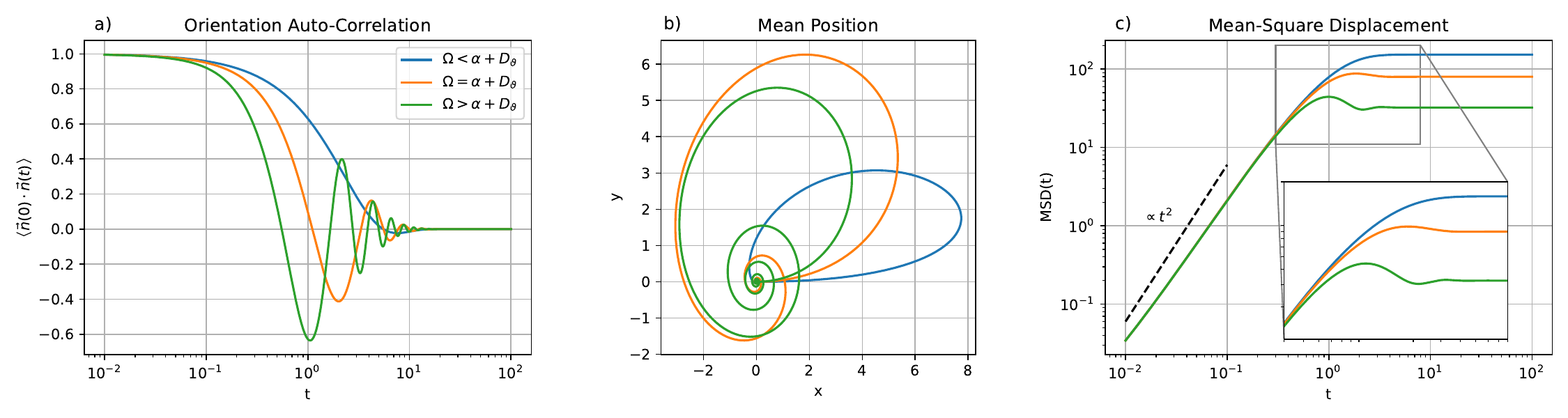}
    \caption{
    Moments of the cABP for different values of the angular velocity $\Omega$ in comparison to the relaxation rate $\lambda=\alpha+D_\vartheta$.
    a) The orientation auto-correlation function 
    $\mean{\vec n(0)\cdot\vec n(t)}$ is a damped oscillation with damping rate $D_\vartheta$. 
    The oscillations occur for $\Omega\geq\lambda=\alpha+D_\vartheta$ and coincide with those in the MSD, cf.~panel c).
    b) The mean position relaxes to the center of the trap. Chirality leads to the almost circular excursions.
    c) The $\text{MSD}(t)$ transitions from a ballistic behaviour $(\propto t^2)$ to a constant steady-state value. Due to chirality the transition shows oscillations 
    for $\Omega\geq\lambda=\alpha+D_\vartheta$. }
    \label{fig:cABP-moments}
\end{figure*}

The time-dependent second moments of the orientation are a means to understand the dynamical behavior of the cABP. 
These quantities are directly related to the typical oscillations observed in the MSD, as illustrated in Fig.~\ref{fig:cABP-moments}. 
The orientation auto-correlation function is the scalar product
\begin{align}
    \mean{\vec n\cdot\vec n'} &= \mean{\cos(\vartheta(t)-\vartheta(t'))}\nonumber\\
    &=\left.\frac{1}{2}\left[\phi^-(t)\phi^+(t')+\phi^+(t)\phi^-(t')\right]\mathcal{Z}\right|_{h,g=0}
\end{align}
with $\vec n=\vec n(\vartheta(t))$ and analogously $\vec n'=\vec n(\vartheta(t'))$. The result is independent of the initial orientation angle $\vartheta_0$, as it must be. We find
\begin{equation}\label{eq:scalar-n-corr}
    \mean{\vec n(t)\cdot\vec n(t')}=\cos\left(\Omega(t-t')\right)\ex^{-D_\vartheta|t-t'|}
\end{equation}
for arbitrary times $t,t'>0$. 
Thus, the orientation auto-correlation function is a damped oscillation with the angular 
frequency $\Omega$ of the particle 
and the time-scale of the exponential decay is given by rotational diffusion, $D_\vartheta$,
cf.~Fig.~\ref{fig:cABP-moments}a). 
This behaviour of the orientation auto-correlation function was reported experimentally   for Janus colloids under optical confinement \cite{Darabi_2026} and  Eq.~(\ref{eq:scalar-n-corr}) had been previously obtained from the FPE by means of a Laplace transform method \cite{BCSmomentsNJP}.

The cross-product of the orientation vector is
\begin{align}\label{eq:cross-n-corr}
    \mean{\vec n\times\vec n'} &= \mean{\sin(\vartheta(t')-\vartheta(t))}\nonumber\\
    &=\left.\frac{1}{2\ii}\left[\phi^-(t')\phi^+(t)-\phi^+(t')\phi^-(t)\right]\mathcal{Z}\right|_{h,g=0}\nonumber\\
    &=\sin\left(\Omega(t'-t)\right)\ex^{-D_\vartheta |t-t'|}.
\end{align}
In 2D this is a pseudo-scalar.
Again, we find a damped oscillation with frequency $\Omega$ and damping rate $D_\vartheta$. For small time differences $\delta t=t'-t\ll1$ this expression reduces to $\mean{\vec n\times\vec n'}\sim \Omega \delta t$, i.e.~a rotation of $\vec n$ at constant angular velocity $\Omega$.

Correlation functions involving the particle position $\vec q(t)$ have to be calculated perturbatively, since the free action only captures the dynamics of a passive particle in harmonic confinement undergoing chiral rotational diffusion. We refer to our previous publication \cite{littek-2025} for the perturbative calculations without chirality and present 
only the results for the mean position and mean-square displacement of the cABP here. 
As already mentioned,
there is a natural truncation of the perturbative series for any moment involving $N$ factors of position 
$\vec q$ at order $\mathcal{O}(S_\mathrm{A}^N)$,
so these results are exact.

For the expectation value of the particle's position and the MSD one obtains
\begin{widetext}
\begin{align}
    \label{eq: mean-pos}
    \mean{\vec q(t)}
    &= \mathrm{e}^{-\alpha t}\vec q_0 + \frac{v_0}{\left(\alpha-D_\vartheta\right)^2+\Omega^2}
    \bigg[\mathrm{e}^{-D_\vartheta t}\left[(\alpha-D_\vartheta)\vec n(t)-\Omega\vec n_\perp(t)\right]-\mathrm{e}^{-\alpha t}\left[(\alpha-D_\vartheta)\vec n(0)-\Omega\vec n_\perp(0)\right]\bigg],\\
    \label{eq:MSD}
    \text{MSD}(t)&=\frac{2D_q}{\alpha}\left(1-\ex^{-2\alpha t}\right)+\frac{v_0^2}{\alpha}\left[\frac{\alpha+D_\vartheta}{(\alpha+D_\vartheta)^2+\Omega^2}+\frac{(\alpha-D_\vartheta)\,\ex^{-2\alpha t}}{(\alpha-D_\vartheta)^2+\Omega^2}\right]-\frac{2v_0^2\,\ex^{-(\alpha+D_\vartheta)t}\cos(\Omega t-\varphi)}{\sqrt{\left[(\alpha-D_\vartheta)^2+\Omega^2\right]\left[(\alpha+D_\vartheta)^2+\Omega^2\right]}}.
\end{align}
\end{widetext}
Here $\vec n_\perp(t) =(-\sin \vartheta(t), \cos\vartheta(t) )^\intercal$. 
The result for $\mean{\vec q(t)}$ states that on average the particle is located in the center of the trap for late times regardless of where the particle started. Fig.~\ref{fig:cABP-moments}~b) shows three trajectories starting at the origin with 
$\vartheta_0=0$ for three different angular velocities $\Omega$. On average, the particle first moves away from the center due to self-propulsion. Then chirality and harmonic confinement bring the particle back to the center. Note that this behaviour of the mean position is also true for the radially peaked distribution,
where the particle orbits (on average) at a finite radius,
see below.

For the MSD, for simplicity we have 
assumed the particle to be in the center of the trap initially, i.e.~$\vec q_0=(0,0)^\intercal$.
For an expression with arbitrary initial conditions $X_0=(\vec q_0,\vartheta_0)$ we refer to Eq.~(94) in Ref.~\cite{littek-2025}. 
The phase angle occurring in the last term 
is given by
\begin{equation}
    \varphi=\arctan\frac{\Omega}{\alpha-D_\vartheta}-\arctan\frac{\Omega}{\alpha+D_\vartheta}.
\end{equation}
One can see that the MSD is a damped oscillation with a finite limit for $t\rightarrow\infty$.

We show the $\text{MSD}(t)$ and 
the orientation auto-correlation function 
$\mean{\vec n(0)\cdot\vec n(t)}$ in Fig.~\ref{fig:cABP-moments}~c) and a), respectively. 
The figure shows the behaviour depending 
on the angular velocity $\Omega$ as compared to the relaxation rate $\lambda=\alpha+D_\vartheta$. 
We want to pay particular attention to the transient dynamics between the ballistic regime and long-time steady-state. The plots show oscillations in the MSD and the orientation auto-correlation with frequency $\Omega$ at intermediate times, as observed experimentally 
in Ref.~\cite{Darabi_2026}. 
The reason for the oscillations is the chiral self-propulsion subject to rotational diffusion and the harmonic confinement. Their interplay leads to a breathing mode of the particle distribution, i.e.~a damped periodic change 
of its width. 

The analytic expression for the MSD allows for a more formal interpretation. The last term in Eq.~(\ref{eq:MSD}) is a damped oscillation with frequency $\Omega$ and relaxation rate 
$\lambda=\alpha + D_\vartheta$. This indicates that there is no sharp transition between an oscillating and non-oscillating transient state depending on the angular velocity and relaxation rate. Rather, the MSD oscillates for {\it any} combination of $\Omega, \alpha$ and $D_\vartheta$. In terms of the time scales we find: For $\Omega<\lambda$ oscillations are suppressed, while for $\Omega>\lambda$ oscillations are only weakly damped. 
One could call this a "breathing crossover". 
It is worthwhile noting that the associated rates $\Omega$ and $\lambda$ are related to typical length-scales of the system: the circular motion of a chiral particle has radius $R=v_0/\Omega$ and the typical spread of the distribution is $l=v_0/\lambda$, such that for $\Omega=\lambda$ these length-scales are the same.

Finally, we calculate the cross correlations of the position and orientation $\mean{q_i(t)n_j(t')}$
with $t'=t+\delta t$. The free contribution is
\begin{align}
    \mean{q_i(t)n_j(t')}^{(0)}&= q_{0i} n_j(t+\delta t)~\ex^{-(\alpha+D_\vartheta)t-D_\vartheta\delta t}.
\end{align}
Since this contribution is purely passive, 
it vanishes for long times $t\rightarrow\infty$ 
due to harmonic confinement and rotational diffusion. 
Activity changes this picture. 
At first order in activity 
(higher order terms $\mathcal{O}(v_0^N)$ with $N\geq2$ 
again vanish identically),
we need to evaluate the integral
\begin{eqnarray}\label{eq: quasi-Green-Kubo}
    \Delta^{(1)}\mean{q_i(t)n_j(t')}
    &=&\ii v_0\int_0^\infty\mathrm{d}s~G_\alpha(t,s)\mean{n_i(s)n_j(t')}\nonumber\\
    &=&v_0\int_0^t\mathrm{d}s~\ex^{-\alpha(t-s)}\mean{n_i(s)n_j(t')},\quad\quad
\end{eqnarray}
which closely resembles a Green-Kubo integral of the velocity autocorrelation with a weight coming from the harmonic trap. 
With the expressions Eqs.~(\ref{eq:scalar-n-corr}, \ref{eq:cross-n-corr}) for the correlation functions of the orientation vector 
we find for the scalar and cross products

\begin{widetext}
\begin{align}
    \Delta^{(1)}\mean{\vec q\cdot\vec n'}&=\frac{v_0\ex^{-D_\vartheta\delta t}}{\sqrt{(\alpha+D_\vartheta)^2+\Omega^2}}\left[\cos(\Omega\delta t+\varphi')-\ex^{-(\alpha+D_\vartheta)t}\cos(\Omega(t+\delta t)+\varphi')\right],\\
    \Delta^{(1)}\mean{\vec q\times\vec n'}&=\frac{v_0\ex^{-D_\vartheta\delta t}}{\sqrt{(\alpha+D_\vartheta)^2+\Omega^2}}\left[\sin(\Omega\delta t+\varphi')-\ex^{-(\alpha+D_\vartheta)t}\sin(\Omega(t+\delta t)+\varphi')\right],
\end{align}
\end{widetext}
where $\varphi'=\arctan\frac{\Omega}{\alpha+D_\vartheta}$. 
For long times, the second term in brackets vanishes. 

These cross correlation functions can be related 
to the diffusivity tensor as pointed out in 
Ref.~\cite{Hargus_2021}. 
Our expression Eq.~(\ref{eq: quasi-Green-Kubo}) 
at equal times, i.e.~$\delta t=0$, is formally 
the Green-Kubo integral evaluated at a finite 
time $t$ for a particle in a harmonic trap. 
Since the system reaches a steady-state at late times, 
we find the associated transport coefficients 
for $t\rightarrow\infty$. Therefore, the isotropic 
(active) diffusivity is
\begin{equation}
    2D_\|=\lim_{t\rightarrow\infty}\mean{\vec q(t)\cdot v_0\vec n(t)}=\frac{v_0^2(\alpha+D_\vartheta)}{(\alpha+D_\vartheta)^2+\Omega^2}.
\end{equation}
The full effective isotropic diffusivity includes another additive term $2D_q$ from the thermal bath. In addition, chirality of the particle gives rise to an odd (active) diffusivity
\begin{equation}2D_\perp=\lim_{t\rightarrow\infty}\mean{\vec q(t)\times v_0\vec n(t)}=\frac{v_0^2\Omega}{(\alpha+D_\vartheta)^2+\Omega^2}.
\end{equation}
Hence, even in the steady-state, there is a non-zero 
flux perpendicular to the gradient of the distribution.
The odd diffusivity is maximal 
for $\Omega=\alpha+D_\vartheta$. 
On the other hand, for $\Omega=0$ the odd diffusivity vanishes, as expected for a non-chiral swimmer.

We note that for the free cABP time-dependent moments in two and three dimensions were also reported in Ref.~\cite{BCSmomentsNJP}, using a Laplace transform of the Fokker-Planck equation to obtain up to the fourth moment of the position and orientation vectors. For $\alpha\rightarrow 0$ our expressions coincide with their results.

\subsection{Probability Distribution}\label{sec:cABP-prob-dist}

As described before, the spatial probability distribution can be calculated in Fourier space from the generating functional as
\begin{align}
    \hat{\mathrm{p}}(t,\vec k) = \mathcal{Z}[\vec h_q =-\ii\vec k\delta _t, 0].
\end{align}
The evaluation at the shifted source $\vec h_q=-\ii\vec k\delta_t$ with $\delta_t=\delta(t-t')$ implies that the perturbative expansion in activity does not terminate naturally. Here, we present the time-dependent probability distribution up to quadratic order in $v_0$. This expansion was shown for the ABP without chirality in Ref.~\cite{littek-2025}. 
For simplicity, we assume initial conditions 
$\vec q_0=(0,0)^\intercal$ and $\vartheta_0=0$. 

The reference motion is the Ornstein-Uhlenbeck process with the well-known probability distribution
\begin{equation}
    \hat{\mathrm{p}}_\mathrm{free}(t,\vec k) = \mathcal{Z}^\alpha[-\ii\vec k\delta_t,0] = \exp\left[-\frac{D_q\vec k^2}{2\alpha}\left(1-\ex^{-2\alpha t}\right)\right]
\end{equation}
given by the stiffness $\alpha$ and the translational diffusion coefficient $D_q$. 

Applying the activity operator $S_\mathrm{A}=-\ii v_0\hat A$ with the definition of Eq.~(\ref{eq: A-operator}), we find a lengthy expression
for the linear correction,
which after inspection turns out to be just
\begin{equation}
    \Delta^{(1)}\hat{\mathrm{p}}(t,\vec k)=\ii\vec k\cdot\mean{\vec q(t)}\hat{\mathrm{p}}_0(t,\vec k)
\end{equation}
for $\vec q_0=(0,0)^\intercal$.
This contribution is purely anisotropic as the self-propulsion shifts the particle position away from the Gaussian reference state. 
Overall, the linear correction to the probability density describes the anisotropic spread of the distribution at early times as the particle self-propels outward. At intermediate times chirality and harmonic confinement inhibit further spread, cf.~Fig.~\ref{fig:cABP-moments} b). Finally, at late times the linear correction is damped to zero by rotational diffusion and harmonic confinement, as expected since the steady-state (late time) distribution is isotropic.

The leading order effect in the long-time limit
is hence contained at quadratic order in $v_0$, 
which is already tough to calculate.
It involves integrals that we abbreviate as
\begin{equation}
    (\sigma\varepsilon):=\int\mathrm{d}t'\int\mathrm{d}t''G_\alpha(t,t')G_\alpha(t,t'')\mathcal{Z}^\vartheta[-\ii\sigma\delta_{t'}-\ii\varepsilon\delta_{t''},0]
\end{equation}
with $G_\alpha$ the Ornstein-Uhlenbeck propagator.
We find the expression
\begin{widetext}
    \begin{equation}
        \Delta^{(2)}\hat{\mathrm{p}}(t,\vec k)=-\frac{v_0^2}{8}\ex^{-2\alpha t}\left\{\vec k^2\left[(+-)+(-+)\right]+(k_x^2-k_y^2)\left[(++)+(--)\right]-2\ii k_xk_y\left[(++)-(--)\right]\right\}\hat{\mathrm{p}}_0(t,\vec k).
    \end{equation}
\end{widetext}
The second order correction involves an isotropic term ($\propto\vec k^2$) and two  anisotropic terms. Solving the $(\sigma\varepsilon)$-integrals 
-- details can be found in Appendix B of Ref.~\cite{littek-2025} for $\Omega =0$ -- 
we find that the anisotropic terms are relevant 
only at intermediate times and vanish at long times. 

Interestingly, the isotropic contribution is again related to a correlation
\begin{equation}
    \Delta^{(2)}\hat{\mathrm{p}}(t,\vec k)=\frac{\vec k^2}{4}\left(\Delta^{(2)}\mean{\vec q^2(t)}\right)\hat{\mathrm{p}}_0(t,\vec k),
\end{equation}
namely to the second order correction to $\mean{\vec q^2(t)}$.
Hence, this term gives rise to an isotropic breathing of the distribution at intermediate times. 
Since the MSD is a constant at late times, 
this is the first perturbative term that
contributes to the steady state distribution. 
Transforming back to configuration space,
the distribution for $t\rightarrow\infty$ 
up to second order in $v_0$ reads
\begin{widetext}
    \begin{equation}\label{eq: Stationary Distribution Path Integral}
    \mathrm{p}^{(2)}_\infty(\vec q)=\frac{\alpha}{2\pi D_q}\left[1-\frac{(\alpha+D_\vartheta)v_0^2}{2D_q[(\alpha+D_\vartheta)^2+\Omega^2]}+\frac{\alpha(\alpha+D_\vartheta)v_0^2}{4D_q^2\left[(\alpha+D_\vartheta)^2+\Omega^2\right]}\vec q^2\right]\ex^{-\frac{\alpha}{2D_q}\vec q^2}.
\end{equation}
\end{widetext}
The shape of the distribution is a modulated Gaussian. Depending on the choice of parameters, the distribution is unimodal with a maximum in the trap center 
or it becomes ring-shaped. We call the transition between these shapes the 'Donut'-transition. 
A first criterion for this shape transition to occur
can be obtained by looking at where 
the curvature of the distribution 
at $q=0$ crosses zero, the unimodal distribution
corresponding to a maximum and
the ring-shape to a minimum in the center.
This leads to a "rule of thumb"
criterion for the velocity
\begin{equation}\label{2nd_order_crit}
    v_0^2 > D_q(\alpha+D_\vartheta)+\frac{D_q\Omega^2}{\alpha+D_\vartheta}.
\end{equation}
For sufficiently large self-propulsion speed $v_0$ the distribution is ring-shaped. As can be seen from the second term, the shape transition is suppressed by chirality, i.e.~chiral particles need to propel themselves at higher velocities compared to non-chiral particles.\\
In the top left panel of Fig.~\ref{fig: cABP Comparison Simulation and Perturbation Theory} we compare our approximate result at order $\mathcal{O}(v_0^2)$ with simulations using $v_0=3$. As expected, the approximation to quadratic order in $v_0$ does not fit the simulations well and even becomes negative in the center of the distribution.

\subsection{Fokker Planck Equation}\label{sec:cABP-FP-equation}

In order to extend the analysis of the stationary probability distribution, Eq.~(\ref{eq: Stationary Distribution Path Integral}), 
beyond the second order $\mathcal{O}(v_0^2)$, 
we can use the power series expansion 
in Fourier space derived from the stationary Fokker-Planck equation in Sec.\ref{secFPE},
\begin{equation}\label{eq:distribution-power-series}
    \displaystyle \hat{P}(\vec{k}) = 2\pi \hat{P}_{m=0}(k) = \left( \sum_{j=0}^{\infty} g_{0,j} k^j \right) \ex^{-\frac{D_q}{2\alpha}k^2}.
\end{equation}

The expansion coefficients $g_{0,j}$ follow 
from the recursion relation, 
Eq.~(\ref{eq:recursion-relation}),
which have the structure
\begin{equation}
 g_{m,j}\propto\big(g_{m-1,j-1}+g_{m+1, j-1}\big).
\end{equation}
Interestingly, this implies a set of
"selection rules" for the coefficients $g_{m,j}$.
Considering $j=0$ first, 
$g_{m,0}$ has to vanish for $m \neq 0$, 
since $g_{m,j}\neq 0$ only for $j\geq0$. 
$g_{0,0}$ is an exception,
since its recursion relation is 
trivially fulfilled and $g_{0,0}=1$ 
has to hold by the normalization  
condition of the  probability density.
For $j>0$, $g_{m,j}$ is determined by the 
coefficients with $m\pm 1$ and $j-1$. 
One can convince oneself that 
if $|m|$ exceeds $j$, 
the respective coefficient vanishes.
Furthermore, 
$j$ and $m$ must always share the same parity (changing  
$j$ by $1$ changes $m$ by exactly $\pm 1$,
hence $m$ changes its parity every time $j$ does). This implies in particular that $g_{0,j}=0$ for all odd $j$, 
reflecting the isotropy of the stationary distribution. In summary, we have found
the rules that  $|m| \le j$ and that 
$ j - |m| $ has to be even.

For the second order ($j=2$), 
Eq.~(\ref{eq:distribution-power-series})
yields exactly Eq.~(\ref{eq: Stationary Distribution Path Integral}), i.e., the same result as calculated via the path integral formalism. We also verified this agreement up to the fourth order by performing the respective path integral calculation in the stationary limit. This agreement in the expansion coefficients stems from the chosen power series ansatz: the Gaussian factor perfectly mirrors the reference motion in the path integral approach.

\begin{figure}[!]
    \centering
    \includegraphics[width=1.0\linewidth]{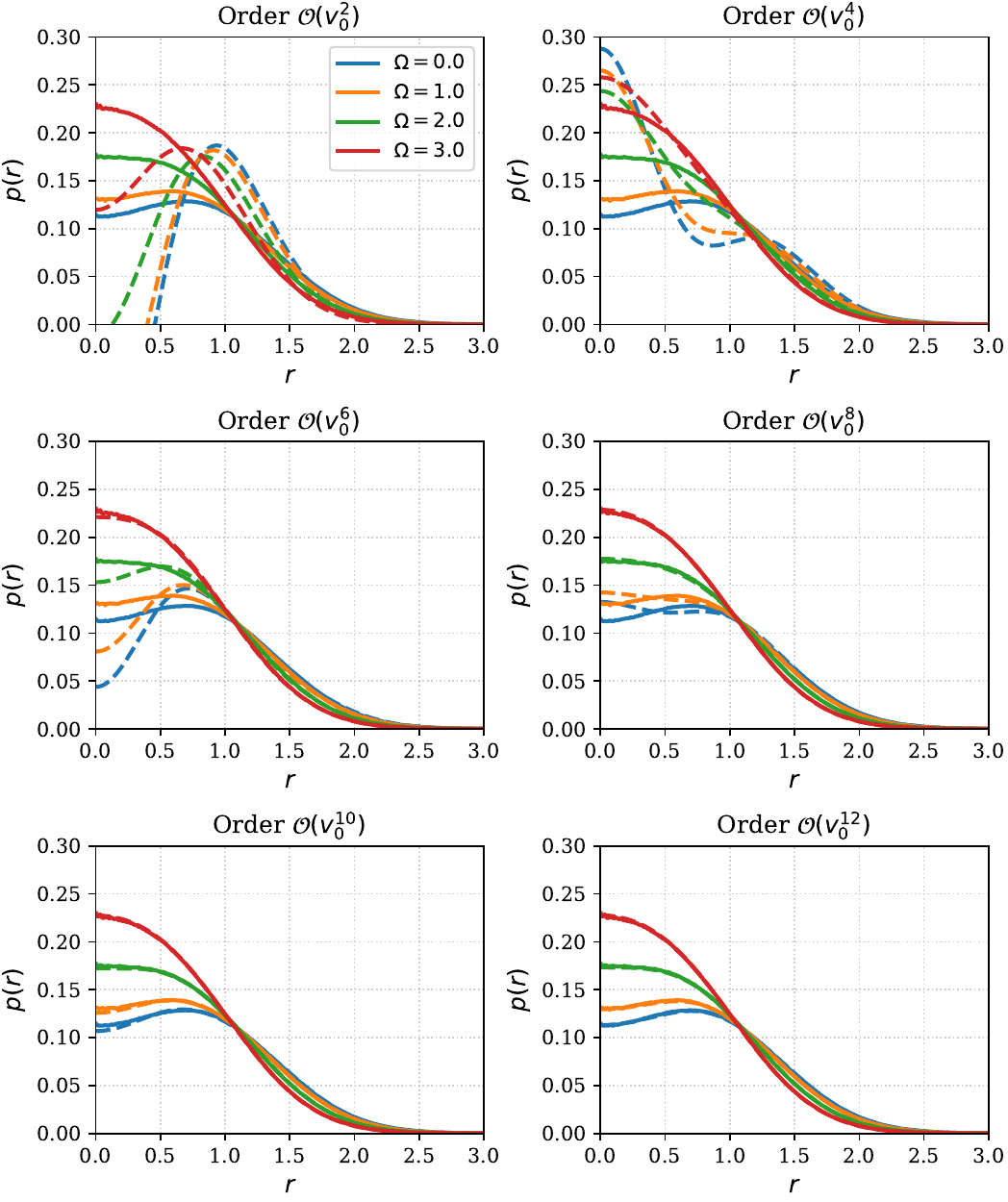}
    \caption{Stationary radial probability distribution of a chiral active particle in a harmonic trap for different values of chirality $\Omega$. The analytical approximation (dashed) converges to Euler-Maruyama simulations (solid) for increasing order $v_0^n$. Parameters: 
    $D_q = 1$, $D_\vartheta = 0.1$, $\alpha = 3$, $v_0 = 3$; simulation time step 
    $\Delta t = 0.001$.}
    \label{fig: cABP Comparison Simulation and Perturbation Theory}
\end{figure}

The inverse Fourier transform 
can be evaluated analytically at all orders. Substituting the variable $c = \frac{r^2}{4a}$, 
one can apply Rodrigues' formula for Laguerre Polynomials $L_n$, see e.g.~\cite{balakrishnan-2020}, 
\begin{equation}
    \frac{\partial^n}{\partial a^n} \left( \frac{1}{a} \ex^{-\frac{c}{a}} \right) = (-1)^n \frac{n!}{a^{n+1}} \ex^{-\frac{c}{a}} L_n\left(\frac{c}{a}\right)
\end{equation}
and the properties of the generating function for the Laguerre polynomials. The inverse Fourier transform then yields
\begin{equation}
    \mathcal{F}^{-1}[k^{2n} \ex^{-a k^2}](\vec{x}) = \frac{n!}{4\pi a^{n+1}} \ex^{-\frac{r^2}{4a}} L_n\left(\frac{\vec{x}^2}{4a}\right), 
\end{equation}
i.e. the $n^{\text{th}}$ order of the perturbative expansion is a closed analytical expression and no numerical Fourier analysis is needed.

The exact expressions are quite cumbersome and therefore not given here. Importantly, in the non-chiral limit ($\Omega=0$), they agree with previously obtained results \cite{malakar-2020,PhysRevLett.129.158001}. 
Fig. \ref{fig: cABP Comparison Simulation and Perturbation Theory} shows a comparison of the radial probability distribution obtained by the power series approximation up to a certain order $v_0^n$ (dashed curves) to Euler-Maruyama simulations (solid curves) of the underlying Langevin equations, for different angular velocities $\Omega$. While the agreement for low orders is poor and the truncated analytical expressions exhibit oscillatory deviations, the approximation converges very well for sufficiently large orders $v_0^n$. 

Looking at the well-converged last panel, we can already conclude that the Donut transition -- i.e.~the distribution exhibiting a maximum at a finite radius $r$ -- is suppressed by chirality: for larger $\Omega$ (red and green curves), the probability density remains peaked at the center of the trap.

\subsection{Phase Diagram}

Using the converged perturbative expansion of the stationary distribution, we can determine the "phase diagram" for the donut transition for arbitrary chirality $\Omega$. This transition was, so far, only investigated for non-chiral particles, either using the same method
\cite{malakar-2020} or by calculating the kurtosis \cite{kurtosis}.
Due to the rotational symmetry of the stationary state, it is again sufficient to evaluate 
$\partial_x^2 \pp(0)$
for a one-dimensional radial section along the $x$-axis. 
The phase diagram can now be obtained by evaluating this criterion across a $(v_0, \alpha)$ parameter grid numerically for a certain order in self-propulsion speed $v_0^n$. 
However, some subtleties should be noted,
namely that the convergence is not "continuous" in the following sense: on the one hand, going up to order $v_0^4$, the donut transition is absent again.
On the other hand, if the perturbation series 
is truncated at any doubly even order 
$n=4m$, $m=2,3,...$, solving for 
zero curvature at the origin 
yields even two critical velocities, $v_0$
and $v_0^*$, for a given value of $\alpha$. 
This suggests a second transition where the particle returns to the confined phase at sufficiently high self-propulsion speeds. 
However, this is just an artifact of the perturbative expansion, since $v_0^*$ 
goes to infinity for increasing order of $n$.

\begin{figure}
    \centering
    \includegraphics[width=0.8\linewidth]{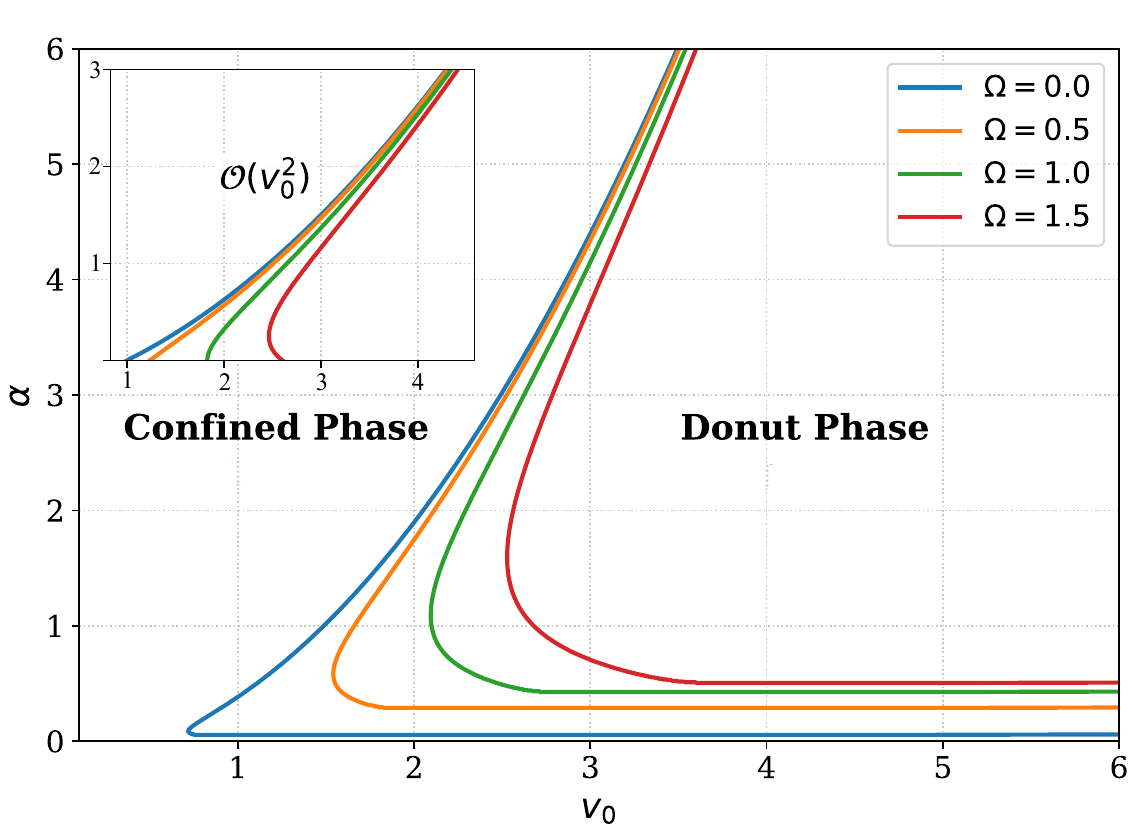}
    \caption{Phase diagram of the donut transition for different values of 
    $\Omega$, obtained at a sufficiently high order such that the analytical expansion fits the Langevin simulations.
    The minimal velocity $v_0$, where a radial peak is possible increases for increasing $\Omega$, indicating that the formation of the "donut-shaped" distribution is suppressed by chirality. The subplot shows the second order approximation, i.e.~the lowest order, corresponding to Eq.~(\ref{2nd_order_crit}).
    }
    \label{fig:donut_phasediag}
\end{figure}

The final phase diagram is shown in 
Fig.~\ref{fig:donut_phasediag}.
It can be seen that the transition is absent for low confinement (low $\alpha$)
and that it needs larger self-propulsion speed $v_0$ for increasing angular
velocity $\Omega$.
The subplot shows the second order approximation, corresponding to Eq.~(\ref{2nd_order_crit}), 
which is 
a nice compact formula, but not accurate.
It can  only recover the behaviour
qualitatively for sufficiently large $\alpha$
and $v_0$.

\subsection{Discussion: comparing the methods}

From the treatment just discussed, the advantages of the two complementary approaches should 
become clear: the path integral approach
is much better suited to calculate moments.
The reason is that the reference motion
-- and its complete time dependence --
are accounted for analytically.
The calculations of moments is a
straightforward field theoretic task
and moreover the moments terminate at
low, finite order. In contrast, 
to obtain the moments from the
Fokker-Planck approach requires
the solution of its full time-dependence,
which is extremely difficult. This 
task was performed (for the non-chiral case) in 
Ref.~\cite{PhysRevLett.129.158001}
and involves complicated combinations
of special functions.
Also note that chirality is just a 
-- rather trivial -- modification of
the reference motion in the path integral
case (namely, the solution gains an additional term $\propto\Omega t$), 
the expansion stays formally the same
with the same propagators. 
In the Fokker-Planck approach, chirality implies a change of eigenfunction structure.

On the flip side, using the path integral
approach for the steady-state
probability distribution is kind of an
overkill: one has to obtain 
the full time-dependent
solution at every order, which becomes
increasingly difficult, and then to
perform the limit towards infinite times.
Here, solving the stationary Fokker-Planck
equation is a much easier task.

But note that, if one is interested
in the (early) time-dependence of the probability
distribution, the path integral 
could again be advantageous: at least the
passive reference motion is accounted
for with its {\it full} time-dependence, 
and one can hence hope to get reasonable approximate
results by accounting for the activity (or interactions between particles) perturbatively.

\section{Chiral Active Brownian Dumbbells}
\label{dumbbell}

\subsection{Model}
To show the versatility of the method,
we now generalize the approach to a
chiral Active Brownian Dumbbell (cABD).
The dumbbell consists of a chiral Active Brownian Particle -- referred to in the following as the swimmer (s) bead -- and a passive Brownian particle -- referred to as the cargo (c) bead -- connected by a harmonic spring of
stiffness $\kappa$. We note that the approach discussed here is also easily generalizable to an active
polymer, composed of more than two particles.

We further consider the dumbbell to be confined in a global harmonic trap of stiffness $\alpha$. Accordingly
the Langevin equations for the two coordinates 
$\vec{q}_s(t), \vec{q}_c(t)$ 
(and in case of the swimmer also its orientation $\vartheta$) read
\begin{equation}\label{eq: Dumbbell equations}
\begin{cases}
    \dot{\vec{q}}_s(t) 
    &= -\alpha \vec{q}_s(t) - \kappa(\vec{q}_s - \vec{q}_c)(t) + v_0 \vec{n}(\vartheta) + \sqrt{2D_s} \vec{\eta}_s(t),\\ \dot{\vartheta}(t) &= \Omega + \sqrt{2D_\vartheta} \eta_\vartheta(t), \\
    \dot{\vec{q}}_c(t) &= -\alpha \vec{q}_c(t) - \kappa(\vec{q}_c - \vec{q}_s)(t) + \sqrt{2D_c} \vec{\eta}_c(t).
\end{cases}
\end{equation}
The noise is again assumed to be Gaussian white noise $\langle \eta_{\mu m}(t) \eta_{\nu n}(t') \rangle = \delta_{\mu\nu} \delta_{mn} \delta(t-t')$, for $m,n \in \{s,c\}$ and $\vartheta$.  If not stated otherwise, 
we assume that the cargo has a smaller diffusion coefficient than the swimmer $D_c\leq D_s$, e.g., due to its large size, but
the opposite case is also interesting and
briefly discussed further below.

Introducing the center of mass $\vec{S}(t)$
and the relative coordinate $\vec{R}(t)$,
\begin{align}\label{eq: Transformation Rule1}
    \vec{S}(t) &= \frac{1}{\sqrt{2}} (\vec{q}_s + \vec{q}_c)(t),\\ 
    \vec{R}(t) &= \frac{1}{\sqrt{2}} (\vec{q}_s - \vec{q}_c)(t), \label{eq: Transformation Rule2}
\end{align}
the deterministic part of the dynamics 
decouples,
\begin{equation}\label{eq: Dumbbell equations in RS-coordinates}
\begin{cases}
    \dot{\vec{S}}(t) &= -\alpha \vec{S}(t) + \frac{v_0}{\sqrt{2}} \vec{n}(\vartheta) + \vec{\xi}_S(t) \\ \dot{\vec{R}}(t) &= -(\alpha + 2\kappa)\vec{R}(t) + \frac{v_0}{\sqrt{2}} \vec{n}(\vartheta) + \vec{\xi}_R(t) \\
    \dot{\vartheta}(t) &= \Omega + \sqrt{2D_\vartheta} \eta_\vartheta(t), 
\end{cases}
\end{equation}
introducing the effective trap stiffnesses 
$\lambda_S=\alpha$ and $\lambda_R=\alpha +2\kappa$. 
From the additivity of variances for independent stochastic processes, the noise for the variables $\vec S$ and $\vec R$ inherits the variance $\langle \xi_{\mu p}(t) \xi_{\nu q}(t') \rangle = 2 \mathcal{D}_{pq} \delta_{\mu\nu} \delta(t-t')$ with the diffusion tensor $\mathcal{D}$ displaying cross-diffusion (off-diagonal terms) as
\begin{equation}
    \mathcal{D}_{pq} = \frac{1}{2} \begin{pmatrix} D_s + D_c & D_s - D_c \\ D_s - D_c & D_s + D_c \end{pmatrix} \equiv \begin{pmatrix} D_+ & D_- \\ D_- & D_+ \end{pmatrix}
\end{equation}
for $p,q\in\{S,R\}$. It also should be noticed 
that the self-propulsion term, $v_0\vec{n}(\vartheta)$, now appears 
in the equations of motion for {\it both} degrees of freedom $\vec S$ and $\vec R$.

As for the cABP, we now apply both the path integral approach and the FPE approach 
to obtain the full,
time-dependent moments, as well as
the steady-state distribution in the trap,
respectively.

\subsection{Generating Functional}

The free generating functional, defined by the passive reference motion for $v_0=0$, is the product
\begin{equation}\label{eq: Generating Functional for cABD}
    \mathcal{Z}_0[h,g]=\mathcal{Z}^\kappa[\vec h_p, \vec g_p]\cdot \mathcal{Z}^\vartheta[h_\vartheta,g_\vartheta],
\end{equation}
where $\vec h_p$ and $\vec g_p$ indicate the source fields for the modes $\vec S$ and $\vec R$,
i.e.~$p\in\{S,R\}$. 
The generating functional for the angle, 
$\mathcal{Z}^\vartheta$,
is the same as before (rotational diffusion). 
The spatial dynamics of the -- passive -- 
dumbbell is captured by $\mathcal{Z}^\kappa$. It also can be written down immediately, 
since the Langevin equations (\ref{eq: Dumbbell equations in RS-coordinates}) define an Ornstein-Uhlenbeck process:
\begin{align}
    \mathcal{Z}^\kappa[\vec{h}_q,\vec{g}_q] &= \exp\left[ 
    \frac{1}{2}\big(\vec{h}\big|C_\kappa\big|\vec{h}\big) + \big(\vec{h}\big|G_\kappa\big|\vec{g}\big) \right],
\end{align}
where we assumed the swimmer and cargo to start at the origin. We also defined a new bracket notation
\begin{equation}
\left( \vec{h} \big| C_\kappa \big| \vec{h} \right) = \sum_{p,q} \int_t \int_{t'} \vec{h}_{p}(t) C_{pq}(t,t') \vec{h}_{q}(t')
\end{equation}
summing over the contributions from $p\in\{S,R\}$ and where a scalar product between vectors is assumed. The bracket notation with the response propagator $G_{\kappa}$ is defined analogously. Finally,
the components of the propagators are
\begin{align}\label{eq: Dumbbell Propagators}
    C_{pq}(t,t')&= 2\mathcal{D}_{pq} \ex^{-\lambda_pt-\lambda_qt} \int_{0}^{\min\{t,t'\}} \dd \tau \, \ex^{(\lambda_p+\lambda_q)\tau},\\
    G_{pq}(t,t')&= -i\Theta(t-t')\ex^{-\lambda_p(t-t')}\delta_{pq} .
\end{align}
As the latter propagator is diagonal, we
introduce $ G_{pq}(t,t')\equiv G_p(t,t')\delta_{pq}$.

The self-propulsion terms induce 
the following contribution in the active part of the action
\begin{equation}
    S_A(\vec \chi,\vartheta) = - \frac{i v_0}{2\sqrt{2}} \sum_p \int \dd t \:\vec{\chi}_{p}(t) \cdot \vec{n}(\vartheta).
\end{equation}
As before, in order to calculate correlation functions order by order in self-propulsion speed $v_0$, 
we deal with $S_A(\vec \chi, \vartheta)$ 
by applying it as an operator, 
exchanging $\vec\chi_{p}$ and $\vartheta$ 
by their functional derivatives. 
This leads to the same structure of the 
activity operator as for the cABP, 
except that now we have to sum over $p=S,R$
\begin{align}
    A &= \frac{1}{2} \sum_p \left\{ \big(a_p^+ | \phi^+\big) + \big(a_p^- | \phi^-\big) \right\} = \sum_p A_p\,.
\end{align}
The action of the operator $a^\sigma_p(t)$ on the passive reference motion
is defined as
\begin{align}
    a^\sigma_p(t)\mathcal{Z}_0[h,g] &=\left[ \big(\vec{h}|G_p\big)_x(t) + \ii\sigma\big(\vec{h}| G_p\big)_y(t) \right] \mathcal{Z}_0[h,g].
\end{align}
Note that, due to $G_{pq}=G_p\delta_{pq}$
one simply has $\big(\vec{h}|G_p\big)_x(t) = \int_{t'} h_{p,x}(t')G_p(t',t)$, where the indices $(x,y)$ refer to the components of $\vec S$ and $\vec R$.

\subsection{Moments}

As before, the moments of $\vec S$ 
and $\vec R$ can be calculated exactly
by applying the activity operator 
to the free generating functional. 
The moments for the positions of the 
particles forming the dumbbell, 
$\vec q_s$ and $\vec q_c$, 
are then obtained by the 
transformation rule, Eqs.~(\ref{eq: Transformation Rule1},
\ref{eq: Transformation Rule2}). 
We will discuss here  
the mean displacement, the 
projections of the relative coordinate
to study dumbbell orientation, 
and the MSD. 
As for the cABP, correlation functions 
of position $\vec q_{s,c}(t)$ of order $N$ can be calculated
exactly, as the expansion in $v_0$ 
terminates at order $v_0^N$.
For simplicity, we choose the 
initial conditions 
$\vec q_s(0)=\vec q_c(0)=0$ and $\vartheta(0)=0$, i.e.,  
swimmer and cargo both start at the origin
(the trap center)
and the initial self-propulsion 
points along the $x$-direction.

\subsubsection{Mean Position}

\begin{figure}
    \centering
    \includegraphics[width=1.0\linewidth]{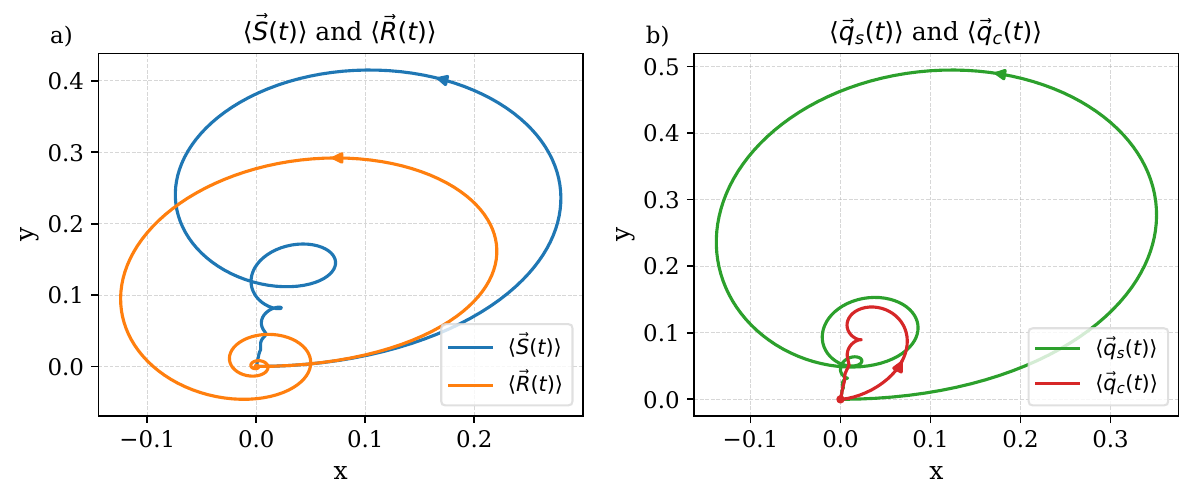}
    \caption{
    a) Trajectories for the mean values 
    of the center of mass vector $\mean{\vec S}$ 
    and the relative vector $\mean{\vec R}$. 
    b) Mean displacement of the swimmer, 
    $\mean{\vec q_s}$, and the cargo, 
    $\mean{\vec q_c}$, constituting the active dumbbell. Parameters: $v_0=1$, 
    $\Omega =2$, $D_\vartheta = 0.5$, $\alpha=0.2$ and $\kappa = 0.3$.}
    \label{fig: Dumbbell Trajectories}
\end{figure}

The resulting mean center of mass coordinate 
$\mean{\vec S(t)}$ and 
relative coordinate $\mean{\vec R(t)}$ 
are identical to the mean displacement 
of the single cABP, cf.~Eq.(\ref{eq: mean-pos})
for the general expression,
but with the trap stiffness $\alpha$ 
replaced by the effective ones,
$\lambda_R$ and $\lambda_S$,
and with a rescaled self-propulsion velocity
\begin{equation}
    \mean{\vec R(t)} = \frac{v_0}{\sqrt{2}}\ex^{-(\alpha+2\kappa) t}\int_0^t\dd\tau \: \vec n(\Omega \tau) \ex^{(\alpha + 2\kappa - D_\vartheta)\tau}. 
\end{equation}
The same expression is valid for $\vec S$,
but with $\kappa =0$.
Here $\vec n(\Omega \tau)=(\cos\Omega\tau,
\sin\Omega\tau)$. 
The mean positions of the two particles,
swimmer and cargo, are then obtained by using
Eqs.~(\ref{eq: Transformation Rule1},
\ref{eq: Transformation Rule2}).
Fig.~\ref{fig: Dumbbell Trajectories} 
shows the mean displacements, allowing to make
several statements: 
Since $\alpha +2\kappa>\alpha$
(and $\kappa>0$),
the mean of the relative mode decays 
faster to zero than the center of mass mode. 
The mean displacement of the swimmer also undergoes 
a larger circle than the cargo. 
Since the swimmer is the one carrying
the activity, it can reach further out 
in the harmonic trap. 
Although the cargo is a passive particle,
due to its coupling to the swimmer
it shows a similar, but smaller 
and more shorter-lived, behavior.  
The excursion of the cargo from the trap center can of course be
increased by increasing the stiffness
$\kappa$ of the connecting spring, 
since the cargo then has to be closer to the swimmer.

\subsubsection{Projections of the Relative Vector}

Since the swimmer is active, while the cargo
is passive, one expects the cargo to 
"trail behind" the swimmer on average. 
This can be analyzed by considering the 
angle between the relative vector 
and the direction of self-propulsion. 
To this end, we define the two (averaged) 
projections parallel and orthogonal to 
the relative vector 
\begin{equation}
    \mean{R^{\parallel}(t)}=\mean{\vec R(t) \cdot\vec n(\vartheta)}\,,\,\,\,
    \mean{R^\perp(t)}=\mean{\vec{R}(t)\times \vec n(\vartheta)}.
\end{equation}
At zeroth order, both projections vanish identically, 
since we have chosen 
$\vec q_s(0) =\vec q_c(0)=(0,0)$.

\begin{figure}
    \centering
    \includegraphics[width=\linewidth]{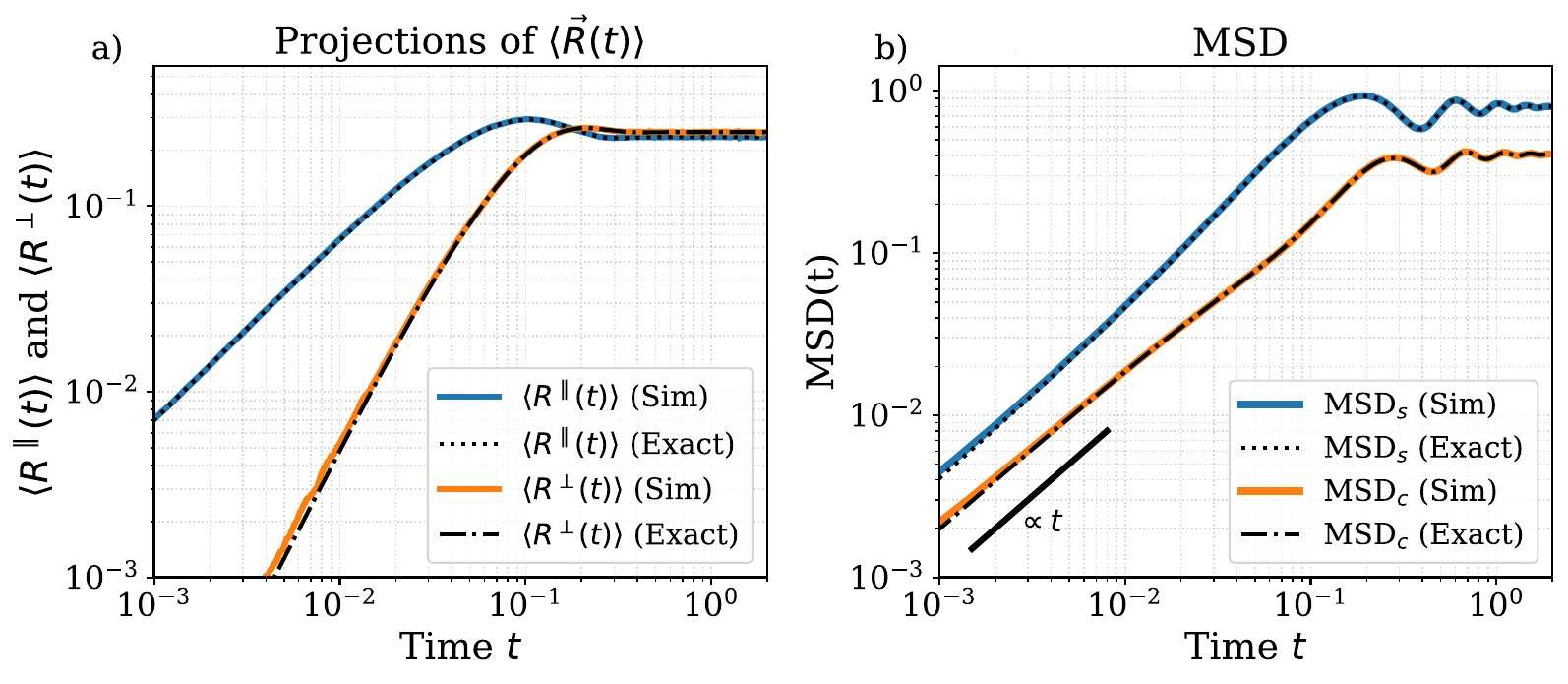}
    \caption{
    \label{fig:mom_dumb}
    Comparison of the analytical expression (solid) and Euler-Maruyama simulations (dashed) of the underlying Langevin equations.
    a) For the projections of the relative vector 
    to the self-propulsion direction, see Eqs.~(\ref{eq: Projections of R, timedependent}), (\ref{eq: Projections of R, timedependent2}). b) For the full 
    time-dependent MSD for the cABD, see Eqs.~(\ref{MSDsinf}), (\ref{MSDcinf})  
    for the stationary expressions. 
    }
    \label{fig: Dumbbell MSD}
\end{figure}

At linear order in $v_0$, we obtain
\begin{widetext}
\begin{align}\label{eq: Projections of R, timedependent}
\mean{R^\parallel(t)} &= \frac{v_0}{\sqrt{2}} \frac{1}{\Omega^2 + (\lambda_1 + D_\theta)^2} \Bigg( (\lambda_1 + D_\theta) + \ex^{-(\lambda_1 + D_\theta)t} \big[\Omega \sin(\Omega t) - (\lambda_p + D_\theta) \cos(\Omega t) \big] \Bigg) \\[1em]
    \mean{R^{\perp}(t)} \label{eq: Projections of R, timedependent2}
    &= \frac{v_0}{\sqrt{2}} \frac{1}{\Omega^2 + (\lambda_1 + D_\theta)^2} \Bigg( \Omega - \ex^{-(\lambda_1 + D_\theta)t} \big[ (\lambda_p + D_\theta) \sin(\Omega t) + \Omega \cos(\Omega t) \big] \Bigg)
\end{align}
\end{widetext}
which is the exact result (all higher orders
vanish again).
Both values show harmonic oscillations with 
angular frequency $\Omega$, similar to 
$\mean{\vec q \times \vec n}$ calculated 
for the cABP. 
Fig.~\ref{fig:mom_dumb}a) shows a comparison
to Langevin simulations.
The asymptotic values for long times are
\begin{align}
    \mean{R^\parallel(\infty)}&=\frac{v_0(\alpha +2\kappa +D_\vartheta)}{\sqrt{2}\big[\Omega^2+(\alpha + 2\kappa +D_\vartheta)^2\big]},\\
    \label{meanRperp}
    \mean{R^\perp(\infty)}&=\frac{v_0\Omega}{\sqrt{2}\big[\Omega^2+(\alpha + 2\kappa +D_\vartheta)^2\big]}.
\end{align}
Since $\mean{R^\parallel(\infty)}>0$, the projection of the self-propulsion onto the average vector is positive, i.e.~relative vector and self-propulsion align. Consequently, the cargo indeed trails
on average behind the swimmer. From Eq.~(\ref{meanRperp}),
chirality introduces a misalignment, as the normal projection $\mean{R^\perp(\infty)}$ is non-vanishing 
for $\Omega\neq0$ (and inherits the sign of $\Omega$). As expected, the chirality of the swimmer particle induces an overall chirality 
of the dumbbell, as also apparent from the mean position and the MSD in $\vec S$. For a proper statement on how much $\vec n$ and $\vec R$ are misaligned, 
however, the
cross-correlation $\mean{R^\|R^\perp}$
has to be evaluated.

\subsubsection{Mean Square Displacement}

The MSD of the swimmer and the cargo
can be calculated from the moments of 
$\vec S$ and $\vec R$ as
\begin{align}\label{MSDs}
    \mathrm{MSD}_s(t)=\langle \vec{q}_s^{\,2}(t) \rangle &= \frac{1}{2} \langle \vec{S}^2 \rangle + \langle \vec{S} \cdot \vec{R}\rangle + \frac{1}{2} \langle \vec{R}^2 \rangle, \\
    \label{MSDc}
    \mathrm{MSD}_c(t)=\langle \vec{q}_c^{\,2}(t) \rangle &= \frac{1}{2} \langle \vec{S}^2 \rangle - \langle \vec{S} \cdot \vec{R}\rangle + \frac{1}{2} \langle \vec{R}^2 \rangle\rangle.
\end{align}
For the passive reference motion, the quadratic correlation functions for $\vec S$ and $\vec R$ can be read off easily,
\begin{align}
    \mean{\vec S^2(t)}^{{(0)}}&=\frac{D_c+D_s}{\alpha}\big(1-\ex^{-2\alpha t}\big),\\
    \mean{\vec R^2(t)}^{{(0)}}&=\frac{D_c+D_s}{\alpha+2\kappa}\big(1-\ex^{-2(\alpha +2\kappa)t}\big),\\
    \mean{\vec S(t)\cdot \vec R(t)}^{{(0)}} &= \frac{D_s-D_c}{\alpha + \kappa}\big(1-\ex^{-2(\alpha+\kappa)t}\big),
\end{align}
as they are the components of the diffusion propagator at equal times, $C_{pq}(t,t)$, see Eq.~(\ref{eq: Dumbbell Propagators}).

The first order correction vanishes for the trivial initial conditions $\vec q_{s,c}(0)=0$ and $\vartheta(0)=0$, since the correction depends linearly on $\vec S(0)$ and $\vec{R}(0)$.
Finally, at second order, 
we have to apply the operator
\begin{equation}
    \hat A^2 = \sum_{p,q}\big(A_p^2 + 2A_p A_q+A_q^2\big)
\end{equation}
to $\mathcal{Z}_0[h,g]$ in Eq.~(\ref{eq: Generating Functional for cABD}). 
The operators $A_p$ and $A_q$ commute for 
different $p,q\in\{S,R\}$, since 
they yield the same expression for both indices
when acting on the free generating functional 
of the angle $\vartheta$.  
With the integrals
\begin{equation}
        (\sigma\varepsilon)_{pq}=\int\mathrm{d}t'\int\mathrm{d}t''G_p(t,t')G_q(t,t'')\mathcal{Z}^\vartheta[-\ii\sigma\delta_{t'}-\ii\varepsilon\delta_{t''},0],
\end{equation}
where $\sigma,\epsilon=\pm1$ and  $p,q\in\{R,S\}$,
the correction of the correlation functions 
at order $v_0^2$ are given by
\begin{align}
    \Delta^{(2)}\mean{\vec S^2}= -\frac{v_0^2}{4}\big\{(+-)_{SS} +(-+)_{SS}\big\}
\end{align}
and similarly for $\Delta^{(2)}\mean{\vec R^2}$. The mixed correlation function correction 
is given by
\begin{equation}
    \Delta^{(2)}\mean{\vec S\cdot\vec R}= -\frac{v_0^2}{2}\big\{(+-)_{SR} +(-+)_{SR}\big\}.
\end{equation}
It should be noted that this non-vanishing 
cross-correlation is not due to different diffusion coefficients, but caused 
by the mixing of the activity operators 
$A_S$ and $A_R$ at second order.

The integrals for the second order expressions are elementary but lead to lengthy expressions.
Therefore, we show only the long-time limits explicitly, which read
\begin{align}
    \langle \vec{S}^2
    \rangle(\infty)&=\frac{D_s+D_c}{\alpha} + \frac{v_0^2 (\alpha + D_\vartheta)}{2\alpha ((\alpha + D_\vartheta)^2 + \Omega^2)},\\
    \langle \vec{R}^2
    \rangle(\infty) &=\frac{D_s+D_c}{\alpha + 2\kappa} + \frac{v_0^2 (\alpha + 2\kappa + D_\vartheta)}{(2\alpha + 4\kappa) ((\alpha + 2\kappa + D_\vartheta)^2 + \Omega^2)}.
\end{align}
Here, one can notice that the relative distance between the cargo and the swimmer, 
$\langle \vec{R}^2(\infty) \rangle$, increases due to the activity, i.e.~the dumbbell spring is stretched compared to the passive case.

In the long-time limit, the cross correlation
and finally the MSDs are given by 
\begin{widetext}
\begin{align}\label{eq: MSD Dumbbells}
    \langle \vec{S} \cdot \vec{R} \rangle (\infty) &= \frac{D_s-D_c}{\alpha + \kappa}
    +
    \frac{v_0^2}{4\alpha + 4 \kappa } \left( \frac{\alpha + D_\vartheta}{(\alpha+D_\vartheta)^2 + \Omega^2} + \frac{\alpha + 2\kappa + D_\vartheta}{(\alpha +2\kappa+D_\vartheta)^2 + \Omega^2} \right),\\[1em]
\label{MSDsinf}
\mathrm{MSD}_s^\infty &= \frac{(D_c + D_s)(\alpha+\kappa)}{\alpha(\alpha+2\kappa)} + \frac{D_s - D_c}{\alpha+\kappa} + \frac{v_0^2}{4(\alpha+\kappa)} \left( \frac{(2\alpha+\kappa)(\alpha + D_\vartheta)}{\alpha \big((\alpha + D_\vartheta)^2 + \Omega^2\big)} + \frac{(2\alpha+3\kappa)(\alpha + 2\kappa + D_\vartheta)}{(\alpha+2\kappa) \big((\alpha + 2\kappa + D_\vartheta)^2 + \Omega^2\big)} \right), \\
\label{MSDcinf}
\mathrm{MSD}_c^\infty &= \frac{(D_c + D_s)(\alpha+\kappa)}{\alpha(\alpha+2\kappa)} - \frac{D_s - D_c}{\alpha+\kappa} + \frac{v_0^2 \kappa}{4(\alpha+\kappa)} \left( \frac{\alpha + D_\vartheta}{\alpha \big((\alpha + D_\vartheta)^2 + \Omega^2\big)} - \frac{\alpha + 2\kappa + D_\vartheta}{(\alpha+2\kappa) \big((\alpha + 2\kappa + D_\vartheta)^2 + \Omega^2\big)} \right),
\end{align}
\end{widetext}
where the latter were obtained by inserting
the asymptotic 
correlations for the $\vec S$,$\vec R$-modes
into Eqs.~(\ref{MSDs}), (\ref{MSDc}).

One can see that the stationary MSD of the swimmer is larger than the one
of the cargo, due to the two minus signs
and all parameters being postive in the analytical expression for $\mathrm{MSD}_c$. 
This again reflects that the activity 
allows the swimmer 
to spread further into the harmonic potential than the passive cargo. 
This fact does not change when increasing $\Omega$, but it leads to an overall decrease
of both long time MSDs.
Furthermore, for $\kappa =0$ and $D_s=D_c$, one regains 
the long time MSD for a cABP for the swimmer 
and the long time result of the 
Ornstein-Uhlenbeck process for the cargo. 

Fig.~\ref{fig: Dumbbell MSD}b)
shows a comparison 
of  Eqs.~(\ref{MSDs}), (\ref{MSDc})
to direct simulations
of the underlying Langevin equations.
Note that the ballistic regime for the given parameters is barely
visible here (in contrast to  Fig.~\ref{fig:cABP-moments}c))
as we focus here on the well-captured oscillations.

\subsection{Fokker Planck Equation}

The path integral calculation for the probability distribution of the chiral Active Brownian Dumbbell (cABD) would be very cumbersome for increasing order of $v_0$. 
As for the cABP in Sec \ref{secFPE}, we hence turn to the stationary FPE now. Again, due to the rotational symmetry of the probability distribution in the long time limit we can perform a mode expansion in Fourier space
and the probability distribution is constructed via a linear recursion relation. For clarity, we reduce the analysis to the case of identical diffusion coefficients $D_q\equiv D_s =D_c$ and comment on the general case later on.

As for the cABP, we start with the Fourier transform of the joint probability for the center of mass mode, the relative mode and the angle
\begin{equation}
    \hat{\PP}(\vec \ell _S, \vec \ell _R, \vartheta) = \int \dd^2 R \, \dd^2 S \, \PP(\vec{R}, \vec{S}, \vartheta) \ex^{-i(\vec{S}\cdot\vec \ell _S + \vec{R}\cdot\vec \ell _
    R)},
\end{equation}
with wave vectors $\vec\ell_{S,R}$. This leads to the stationary FPE in Fourier space
\begin{widetext}
\begin{align}
    0 &= -\alpha \vec{\ell}_S \cdot \nabla_{\ell_s} \hat{\PP} - (\alpha + 2\kappa) \vec{\ell}_R \cdot \nabla_{\ell_R} \hat{\PP} -D_q(\vec \ell_S^2 + \vec{\ell}_R^2) \hat{\PP} - \Omega \partial_\vartheta \hat{\PP} + D_\vartheta \partial_\vartheta^2 \hat{\PP} - \frac{iv_0}{\sqrt{2}}\vec{n}(\vartheta)\cdot (\vec{\ell}_S + \vec{\ell}_R) \hat{\PP}.
\end{align}
\end{widetext}
In order to use the rational symmetry of the stationary distribution, we decompose the wave vectors in polar coordinates
\begin{equation}
    \vec{\ell}_{S,R} = \ell_{S,R} \begin{pmatrix} \cos\varphi_{S,R} \\ \sin\varphi_{S,R} \end{pmatrix}.
\end{equation}
The coupling terms to the self-propulsion can then be written as
\begin{align}
    \vec{n}(\vartheta) \cdot \vec{\ell}_{S,R} = \frac{\ell_{S,R}}{2} \left( \ex^{i(\vartheta-\varphi_{S,R})} + \ex^{-i(\vartheta-\varphi_{S,R})} \right).
\end{align}
In contrast to the cABP, we now have to perform a mode decomposition in two angles, 
\begin{align}
    \hat{\PP}(\vec{\ell}_S, \vec{\ell}_R, \vartheta) = \hspace{-5mm}\sum_{m_S, m_R=0}^\infty\hspace{-3mm} \hat{\PP}_{m_S, m_R}(\ell_S, \ell_R) \ex^{-im_S(\vartheta-\varphi_S) - im_R(\vartheta-\varphi_R)}
\end{align}
and the mode coefficients are of the form
\begin{equation}
    \hat\PP_{m_S,m_R}(\ell_S,\ell_R)= \Bigg(\sum_{j_S,j_R=0}^\infty g_{m_S,m_R}^{j_S,j_R} \ell_S^{j_S}\ell_R^{j_R}\Bigg) \hat \PP_{0,0}(\ell_S,\ell_R).
\end{equation}
The passive and isotropic state $\hat\PP_{(0,0)}$ is obtained in a similar way as for the cABP and reads
\begin{align}\label{passive_cABD}
    \hat\PP_{0,0}(\ell_S,\ell_R)= \ex^{-\frac{D_q}{2\alpha}\ell_S^2 - \frac{D_q}{2(\alpha + 2\kappa)}\ell_R^2}.
\end{align}
From the stationary FPE, we get 
the recursion relation 
for the coefficients of the mode expansion 
\begin{widetext}
\begin{align}
    \big[\alpha j_S + (\alpha+2\kappa)j_R + D_\vartheta(m_S+m_R)^2 - i(m_S+m_R)\Omega\big] g_{m_S m_R}^{j_S j_R} 
    = \frac{-i v_0}{2\sqrt{2}} \big[ g_{m_S-1, m_R}^{j_S-1, j_R} + g_{m_S+1, m_R}^{j_S-1, j_R} + g_{m_S, m_R-1}^{j_S, j_R-1} + g_{m_S, m_R+1}^{j_S, j_R-1} \big],
\end{align}
\end{widetext}
which is a perturbative expansion in orders $v_0^{j_S+j_R}$ of the probability densities in Fourier space for the center mode $\vec S$ and the relative mode $\vec R$.

In order to get the probability densities for the swimmer and cargo positions, we change the coordinates of the isotropic $(m_S,m_R)=(0,0)$-mode from $(\vec S,\vec R)$ to $(\vec q_s,\vec q_ c)$. Since this is an orthogonal transformation, the Jacobian is trivial and we get
\begin{equation}
 \hat{\PP}_{0,0}(\vec{k}_s, \vec{k}_c)
 =\hat{\PP}_{0,0}\left(
 \vec{\ell}_S = \frac{\vec{k}_s+\vec{k}_c}{\sqrt{2}}, 
 \vec{\ell}_R = \frac{\vec{k}_s-\vec{k}_c}{\sqrt{2}}\right).
\end{equation}
By setting $\vec k_s$ or $\vec k_c$ to zero, we get the Fourier transform of the distribution for $\vec q_c$ or $\vec q_s$ respectively.

\begin{figure}[b!]
    \centering
    \includegraphics[width=0.99\linewidth]{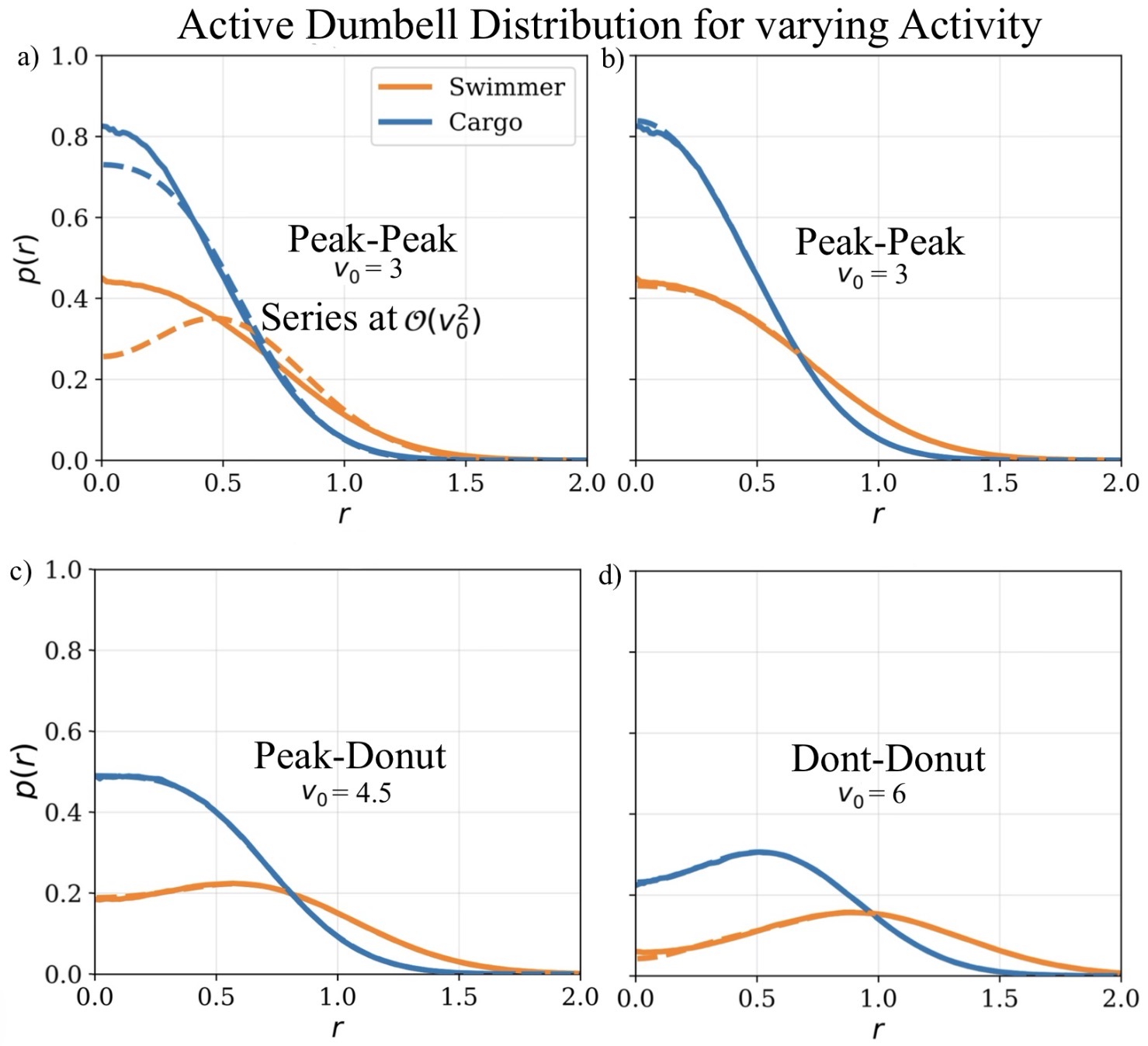}
    \caption{Comparison of the analytical expressions (dashed) and Euler-Maruyama simulations (solid) of the probability distribution of a harmonically confined 
    chiral active Brownian dumbbell 
    for different values of activity, i.e.~propulsion speed $v_0$.
    The distribution for the swimmer bead is 
    shown in orange and the cargo
    bead in blue. The values of activity are chosen such that the three different qualitative regimes are apparent, cf.~also the phase diagram in Fig.~\ref{fig:Phasediagram Dumbbell}. As a reference, a) shows the second order of the perturbation theory for $v_0=3$. In the other plots, the order of the expansion is chosen such that the analytical expansion is close the the simulation. Parameter values are $\kappa = 6$, $\Omega = 3$, $D_s = 1$, $D_c = 0.5$, $D_\vartheta = 0.1$.}
    \label{fig: Dumbbell Distributions}
\end{figure}

Similar to the cABP, the distribution constructed from the recursion relation converges to the simulation, as shown in  Fig.~\ref{fig: Dumbbell Distributions}a) vs.~b). Interestingly,
a series of donut transitions can now be observed:
for small self-propulsion velocity,
both the swimmer (orange) and the cargo (blue)
are peaked in the trap center,
see Fig.~\ref{fig: Dumbbell Distributions}b).
Increasing velocity beyond a threshold, 
first the swimmer enters
the radially peaked donut phase, 
cf.~Fig.~\ref{fig: Dumbbell Distributions}c),
because it is the
active part of the dumbbell. Only if a second
velocity threshold is passed, also the passive
cargo feels or "inherits" the activity and
also becomes radially peaked, see 
Fig.~\ref{fig: Dumbbell Distributions}d).

\begin{figure*}
    \centering
    \includegraphics[width=\linewidth]{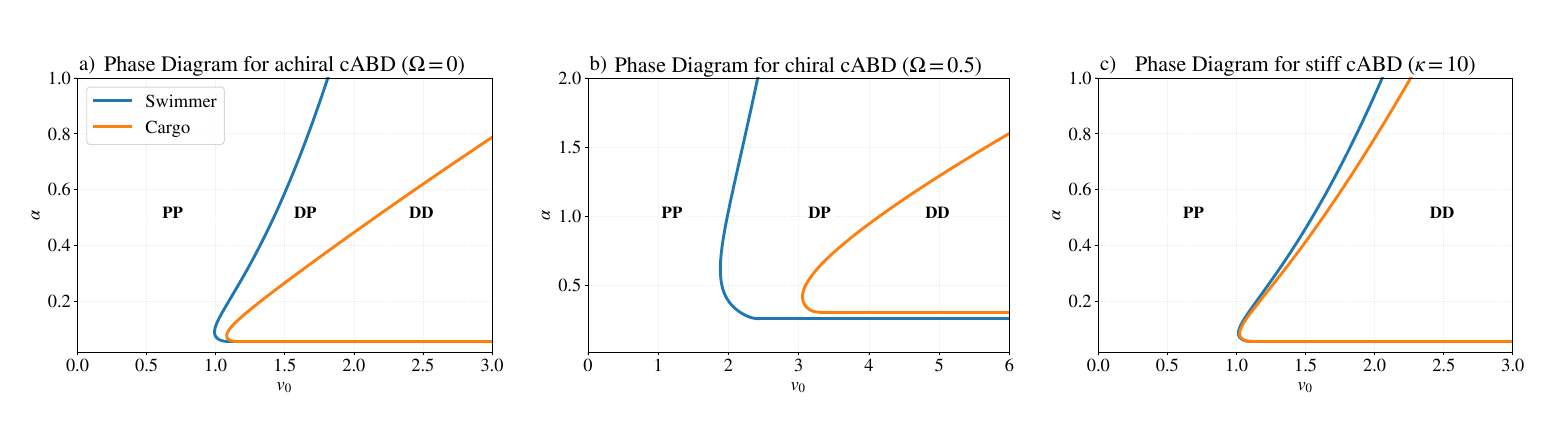}
    \caption{Phase diagrams marking the donut transitions for the swimmer and the
    cargo of a cABD, respectively.
    In all panels, a sufficiently high order ($\mathcal{O}(v_0^{24})$) was used, such that the analytical expansion fits the Langevin simulations.
    Panel a) shows the case of an achiral ABD and the three different phases PP (peak-peak, i.e.~both peaked), DP (swimmer in donut phase, cargo peaked) and DD 
    (both swimmer and cargo in donut phase). 
    The chosen parameter values are $D_q=1.0$, $D_\vartheta=0.1$, $\kappa =1$ and $\Omega =0$. Panel b) shows the chiral case, with angular velocity $\Omega =0.5$ and otherwise unchanged parameters. Both phase boundaries shift to larger velocities and the DP-phase region is increased. Panel c) shows the case of a very stiff achiral dumbbell, where $\kappa =10$. The DP phase shrinks, showing that the dumbbell behaves more and more like a single cABP upon increasing the spring stiffness.}
    \label{fig:Phasediagram Dumbbell}
\end{figure*}

In Fig.\ref{fig:Phasediagram Dumbbell}
we constructed the phase diagram for the swimmer's and cargo's distribution,
both for the achiral case in a) and the chiral one in b).
It should be compared to the diagram
for the cABP in Fig.~\ref{fig:donut_phasediag}.
The donut transition was again obtained via
the sign change of the second derivative at the origin. As already evident from
Fig.~\ref{fig: Dumbbell Distributions}, 
there now are three different phases: 
at small velocities, both the swimmer and the cargo 
are trap-centered. At a first threshold velocity of the swimmer, the swimmer enters the donut phase, but the cargo does not, as it is a passive particle and the inherited activity from the swimmer is 
not sufficiently high. Only at large velocities, both particles are in the donut phase.

Increasing the spring stiffness $\kappa$, 
cf.~Fig.\ref{fig:Phasediagram Dumbbell}c),
the mixed Peak-Donut-phase becomes narrower 
and the two phase boundaries finally converge. In this rigid limit, 
the dumbbell effectively behaves 
like a single cABP, as it should. 

Let us now comment on the general case,
where the diffusion coefficients for swimmer and cargo are not equal. There are two significant changes: First, the passive reference state, cf.~Eq.(\ref{passive_cABD}), is no longer a pure Gaussian, but contains a factor of a modified Bessel function
\begin{align}
    \hat \PP_{0,0}(\ell_S,\ell_R)=I_0\big(\tfrac{D_s-D_c}{2(\alpha + \kappa)}\ell_S\ell_R\big)\  \ex^{-\frac{D_s+D_c}{4\alpha}\ell_S^2-\frac{D_s+D_c}{4(\alpha + 2\alpha)}\ell_R^2}.
\end{align}
Second, the recursion relation for the coefficients  $g_{m_S,m_R}^{g_S,g_R}$ contains an additional term, which accounts for the cross diffusion of swimmer and cargo fluctuations in the $(\vec S,\vec R)$-system. As a consequence, the selection rules for the expansion change such that $j_s+j_c$ does no longer agree with the expansion order in $v_0$. This could be remedied by a slight change of the mode expansion, but is judged beyond the scope of this work.

Instead we comment here 
on another interesting effect that can be
discussed at the level of the  already obtained results for the moments. Note that in sections A-C we assumed that the cargo has a smaller translational diffusion coefficient than the swimmer. This reflects the classical example of, e.g., molecular
motors carrying a large cargo such as a vesicle. An active Janus particle 
could however also carry a smaller passive particle connected via flexible polymer linker(s). In that case, $D_c$ could  also be larger than $D_s$. Interestingly, there is a threshold value of $D_c$ where the MSD of the cargo becomes larger than that of the swimmer. It is given by the condition 
\begin{equation}
D^*_c=D_s+\frac{v_0^2}{2}\Bigg[\frac{\alpha + D_\vartheta}{(\alpha +D_\vartheta)^2+\Omega^2}+\frac{\alpha + 2\kappa +D_\vartheta}{(\alpha+2\kappa +D_\vartheta)^2+\Omega^2}\Bigg],
\end{equation}
as can be read off Eqs.~(\ref{MSDsinf}), (\ref{MSDcinf}). Although the parallel component of the relative vector along the self-propulsion does not change (since it does not depend on the translational diffusion), the motion of the dumbbell is now dominated by the strong diffusion of the cargo and it hence can spread further than the swimmer. 
Interestingly, this effect can be so strong, that the distribution of the swimmer
-- which is in the donut phase, i.e.~radially peaked, for small cargo diffusion -- 
becomes peaked at the trap center again for
large cargo diffusion. 
The cargo hence dominates the fluctuating dynamics of the system for $D_c>D_c^*$, 
and one could call this
a "Tail-wags-the-Dog" transition: metaphorically, if the tail of a dog wags too strongly, the dog is essentially wagged by its tail. 

For increasing chirality $\Omega$ the difference between $D_s$ and $D_c^*$ becomes smaller, since the swimmer moves on closer and closer circles and a smaller 
diffusion coefficient of the cargo is needed in order to dominate the diffusive motion. 

\section{Conclusions and outlook}
\label{sec:conclusions}

We have demonstrated that the combination
of the MSRJD path integral formalism
and the Fokker-Planck approach constitutes
a versatile tool to analyze active Brownian
particles, including chiral ones and 
dumbbells with an active and a passive bead,
trapped in a harmonic potential in an exhaustive
manner. The MSRJD method is best suited
to evaluate time-dependent moments.
It leads to exact results,
by performing a perturbation in activity/velocity
which terminates at the order of the moment.
The FPE approach is best suited
to obtain the steady state probability distribution, as
it leads to a recursion relation for a
series expression that is easy to evaluate
analytically to arbitrary order.

We first applied the method 
to the chiral active Brownian particle (cABP)
in a harmonic trap.
We obtained the moments of both position
(up to MSD) and orientation exactly, as well as
the cross correlation, which could be related
to odd diffusion. Using the FPE approach,
we obtained the steady state probability
distribution as a series expansion.
Comparing against a numerical solution
of the underlying Langevin equations,
we showed convergence and obtained the "phase 
diagram" of the transition from a trap-centered
distribution
to a ring-shaped "donut" distribution.
We show that chirality renders this transition more difficult,
i.e.~a higher self-propulsion velocity $v_0$
is needed. This generalizes analytical
results obtained previously for trapped
non-chiral particles  \cite{malakar-2020,kurtosis}.

We then applied the combined approach to the
trapped active Brownian dumbbell (cABD),
which to our knowledge has not been studied so far. We obtained that the MSD
is larger for the active ("swimmer") bead than
for the passive ("cargo") bead.
On average, the spring between the two particles is stretched compared to the fully passive dumbbell.
The relative vector and the direction of self-propulsion are on average aligned, i.e.~the cargo 
trails behind the swimmer,
an effect diminished by chirality.
In case the active particle is chiral,
this chirality is inherited by the passive
bead as well.
Concerning the transition in the trap,
we found a new series of transitions,
where first the active bead transitions to
a ring-shaped distribution and  
the passive bead transitions
only for higher activity. Finally, we found
that for a cargo with a diffusion
coefficient higher than the one of the 
active bead,
this transition can be reverted, i.e.~the
fluctuations of the passive bead dominate
the behaviour and the active particle
becomes trap-centered again.

In the future, it will be interesting
to generalize the approach to longer active
polymers, also including ring-shaped polymers and different boundary conditions. 
It should be noted that,
similar to here by introducing the 
center of mass and
relative coordinate,
introducing the concept of normal modes,
the Rouse chain dynamics can be diagonalized
and hence the MSRJD path integral can be applied
in exactly the same way.

Other interesting questions relate to
generalizing the potential to non-harmonic
contributions, and to study interactions
of particles. These extensions are much harder
-- in the MSRJD approach they require additional
perturbation series beyond activity
and in the FPE approach, they induce 
nonlinearities -- but nevertheless constitute
a route worth trying to shed new light on
fundamental active matter systems.

\section*{Acknowledgments}
We acknowledge support from the Deutsche Forschungsgemeinschaft (DFG, German Research Foundation) under Germany’s Excellence Strategy EXC 2181/1–390900948 (the Heidelberg STRUCTURES Excellence Cluster) and from the European Union through the MSCA Doctoral Network CAFE-BIO (101226762). 

\bibliography{References}

\end{document}